\documentclass[twocolumn,usenames,dvipsnames]{aastex63}

\newcommand\al{$\alpha$}
\newcommand\dnu{$\Delta\nu$}
\newcommand\nmax{$\nu_{\mathrm{max}}$}
\newcommand\Kepler{\textit{Kepler }}
\newcommand\ktwo{\textit{K2 }}

\definecolor{scarlet}{HTML}{bb0000}
\definecolor{blubla}{HTML}{23B5D3}
\definecolor{stellocommentstoaddress}{HTML}{A2AD91}

\usepackage [english]{babel}
\usepackage [autostyle, english = american]{csquotes}
\MakeOuterQuote{"}

\received{}
\revised{}
\accepted{}

\submitjournal{AJ}

\shorttitle{Intermediate Age Population}
\shortauthors{Warfield et al.}

\begin{document}

\title{An Intermediate-age Alpha-Rich Galactic Population in \textit{K2}}

\email{jtw5zc@virginia.edu}

\author{Jack T. Warfield}
\affiliation{Department of Astronomy, The Ohio State University, 140 West
18th Avenue, Columbus, OH 43210, USA}
\affiliation{Department of Physics,
The Ohio State University, 191 West
Woodruff Avenue, Columbus, OH 43210, USA}
\affiliation{Department of Astronomy, The University of Virginia, 530 McCormick Road, Charlottesville, VA, 22904, USA}

\author{Joel C. Zinn}
\affiliation{School of Physics, University of New South Wales, Barker                           
  Street, Sydney, NSW 2052, Australia}
\affiliation{Department of Astronomy, The Ohio State University, 140 West                        
  18th Avenue, Columbus, OH 43210, USA}

\author{Marc H. Pinsonneault}
\affiliation{Department of Astronomy, The Ohio State University, 140 West 18th Avenue, Columbus, OH 43210, USA}

\author{Jennifer A. Johnson}
\affiliation{Department of Astronomy, The Ohio State University, 140 West                        
  18th Avenue, Columbus, OH 43210, USA}
\affiliation{Center for Cosmology and AstroParticle Physics, 
The Ohio State University          
Columbus, OH 43210, USA}

\author{Dennis Stello}
\affiliation{School of Physics, University of New South Wales, Barker
  Street, Sydney, NSW 2052, Australia}
\affiliation{Sydney Institute for Astronomy (SIfA), School of Physics,
  University of Sydney, NSW 2006, Australia}
\affiliation{Stellar Astrophysics Centre, Department of Physics and
Astronomy, Aarhus University, Ny Munkegade 120, DK-8000
Aarhus C, Denmark}
\affiliation{Center of Excellence for Astrophysics in Three Dimensions
(ASTRO-3D), Australia}
\author{Yvonne Elsworth}
\affiliation{School of Physics and Astronomy, University of Birmingham, Edgbaston, Birmingham, B15 2TT, UK}
\affiliation{Stellar Astrophysics Centre, Department of Physics and Astronomy, Aarhus University, Ny Munkegade 120, DK-8000 Aarhus C, Denmark}
\author{Rafael A. Garc{\'i}a}
\affiliation{IRFU, CEA, Universit{\'e} Paris-Saclay, F-91191 Gif-sur-Yvette, France}
\affiliation{AIM, CEA, CNRS, Universit{\'e} Paris-Saclay, Universit{\'e} Paris Diderot, Sorbonne Paris Cit{\'e}, F-91191 Gif-sur-Yvette, France}
\author{Thomas Kallinger}
\affiliation{Institute of Astrophysics, University of Vienna, T{\"u}rkenschanzstrasse 17, Vienna 1180, Austria}
\author{Savita Mathur}
\affiliation{Space Science Institute, 4750 Walnut Street Suite \#205, Boulder, CO 80301, USA}
\affiliation{Instituto de Astrof\'{i}sica de Canarias (IAC), E-38205 La Laguna, Tenerife, Spain}
\affiliation{Departamento de Astrof\'{i}sica, Universidad de La Laguna (ULL), E-38206 La Laguna, Tenerife, Spain}
\author{Beno{\^i}t Mosser}
\affiliation{LESIA, Observatoire de Paris, PSL Research University, CNRS, Sorbonne Universit{\'e}, Universit{\'e} de Paris Diderot, 92195 Meudon, France}

\author{Rachael L. Beaton}
\altaffiliation{Hubble Fellow\\Carnegie-Princeton Fellow}
\affiliation{Department of Astrophysical Sciences, Princeton University, 4 Ivy Lane, Princeton, NJ 08544, USA}

\author{D. A. Garc\'{i}a-Hern\'{a}ndez}
\affiliation{Instituto de Astrof\'{i}sica de Canarias (IAC), E-38205 La Laguna, Tenerife, Spain}
\affiliation{Departamento de Astrof\'{i}sica, Universidad de La Laguna (ULL), E-38206 La Laguna, Tenerife, Spain}

\begin{abstract}

We explore the relationships between the chemistry, ages, and locations of stars in the Galaxy using asteroseismic data from the \ktwo mission and spectroscopic data from the Apache Point Galactic Evolution Experiment survey. Previous studies have used giant stars in the \textit{Kepler} field to map the relationship between the chemical composition and the ages of stars at the solar circle. Consistent with prior work, we find that stars with high [\al/Fe] have distinct, older ages in comparison to stars with low [\al/Fe].
We provide age estimates for red giant branch (RGB) stars in the \textit{Kepler} field, which support and build upon previous age estimates by taking into account the effect of \al-enrichment on opacity. Including this effect for [\al/Fe]-rich stars results in up to 10\% older ages for low-mass stars relative to corrected solar mixture calculations. This is a significant effect that Galactic archaeology studies should take into account.
Looking beyond the \Kepler field, we estimate ages for 735 red giant branch stars from the \ktwo mission, mapping age trends as a function of the line of sight. We find that the age distributions for low- and high-[\al/Fe] stars converge with increasing distance from the Galactic plane, in agreement with suggestions from earlier work. We find that \ktwo stars with high [\al/Fe] appear to be younger than their counterparts in the \textit{Kepler} field, overlapping more significantly with a similarly aged low-[\al/Fe] population.
This observation may suggest that star formation or radial migration proceeds unevenly in the Galaxy.

\end{abstract}


\section{Introduction} \label{sec:intro}

The study of the chemical evolution of the Milky Way has a rich history and a vibrant present. Historically, it has been easier to measure abundances than it has been to measure ages. As a result, most studies have relied on abundance data alone---for example, very low absolute iron abundance, or characteristic heavy element abundance patterns relative to iron---as indicators of membership in old populations. However, we can now measure ages for large samples of evolved stars using stellar pulsation---or asteroseismic---data from large time domain surveys, and we have an unprecedented wealth of abundance data from massive spectroscopic surveys. These new tools allow us to trace out the enrichment history of the Milky Way in a far more detailed fashion than was possible even a few years ago. In this paper we focus on ages for evolved red giants with enhanced abundances of $\alpha$-capture elements relative to iron compared to the Sun, which we will refer to as $\alpha$-rich stars.

The existence of $\alpha$-rich stars has been known for over half a century \citep{aller1960,wallerstein1962}. 
The $\alpha$-capture elements---such as O, Mg, Si, S, Ca, and Ti---are primarily produced in core-collapse supernovae of massive, short-lived stars (SNe II). Fe-peak elements, by contrast, can also be injected into interstellar medium by the explosive destruction of a white dwarf (a Type Ia supernova, SNe Ia). The latter process requires a longer-lived progenitor.
Therefore, in very old populations, such as the Galactic halo, stars are thought to be $\alpha$-rich because they formed before SNe Ia occurred in significant numbers \citep{Tinsley1979}.

In a simple chemical evolution model, the [$\alpha$/Fe] ratio of the gas would decline as the number of SNe Ia contributing Fe-peak elements increases, before reaching an equilibrium ratio \citep[e.g.,][]{Weinberg+2017}.  This is not what is observed in the solar neighborhood \citep{prochaska2000,Bensby2003}. Rather than a single sequence from high-$\alpha$ to low-$\alpha$, stars with $-1 <$ [Fe/H] $< 0$ can have a range of [$\alpha$/Fe] values. 
The origin of this bi-modal $\alpha$ sequence of stars in the Galaxy is still a puzzle. \cite{Bensby2003} showed that there are distinct trends in [$\alpha$/Fe] vs. [Fe/H] space for the geometrically defined thin and thick discs \citep{GilmoreReid1983}. \cite{Hayden2015} investigated how these trends behave as a function of Galactic radius ($R$) and height above the Galactic plane ($Z$) using $\sim 70,000$ red giants from the Sloan Digital Sky Survey (SDSS) Apache Point Galactic Evolution Experiment (APOGEE) Data Release (DR) 12. These authors found the high-$\alpha$ part of this sequence strongly present only at $|Z| > 0.5$ kpc and $R < 11$ kpc.

Because of the origins of the \al\ elements, the \al-rich and \al-poor populations must be somehow tied to the history of SNe II and Ia. However, all of the observed patterns---for instance, the spread in [\al/Fe] at a given [Fe/H]---cannot be explained by the SNe history alone. Therefore, other, more complex mechanisms must also contribute to the formation of the observed chemical and age patterns.
Some proposed mechanisms capable of reproducing the observed sequences are the radial migration of stars in the Galaxy \citep[see, e.g.,][]{sellwoodbinney2002,SB2009,NBB2014,Weinberg+2017,sharma+2020b}; two separate star formation episodes driven by pristine gas infall into the Galaxy \citep[e.g.][]{chiappini_matteucci_gratton1997, Spitoni2019, lian+2020}; and stars forming throughout the Galaxy in clumpy bursts \citep[e.g.][]{Clarke2019}.
Though all are able to roughly recreate the observed chemical pattern, each of these mechanisms are responsible for different predictions concerning the relative ages of the \al-rich and \al-poor populations and how homogeneous these results will be across the Galaxy. Therefore, finding ages for stars as functions of their chemical abundances and locations in the Galaxy is an important step in uncovering the mechanisms that are responsible for the Galaxy's formation.

Finding ages for large numbers of field red giants up to several kpc away from the Sun is now possible through asteroseismology. Near-surface turbulent convective motions in cool stars generate sound waves, and for distinct resonant frequencies of the stellar interior, standing waves can be induced, forming a frequency pattern of overtone modes of different spherical degrees. The frequency spacings between the radial modes (\dnu) is related to the mean density \citep{ulrich1986,kjeldsen&bedding1995}. The frequency of maximum acoustic power (\nmax) is related to the acoustic cut-off frequency, and therefore the density scale height and the surface gravity \citep{brown1991,kjeldsen&bedding1995,chaplin+2008}. It is therefore possible to solve for mass and radius through scaling relations if \dnu, \nmax, and $T_{\rm eff}$ are known. \cite{apokasc1, apokasc2} produced catalogs of values for \nmax, \dnu, asteroseismic masses \& radii, chemical abundance estimates, and age estimates for targets in the fields observed photometrically by NASA's \Kepler mission and spectroscopically by APOGEE.

Ages for the \al-rich and \al-poor populations in the \cite{apokasc1} data set---including sample selection effect corrections---were calculated by \cite{SilvaAguirre2018} (hereafter SA18). These ages were found using different asteroseismic corrections than used for the ages published by \cite{apokasc2}. Partially motivated by finding chemical signatures that can be used as tracers of the thin and thick disks, SA18 used a combination of photometric, spectroscopic, and asteroseismic parameters to estimate the ages of 1590 red giant branch and red clump stars located within the \Kepler field. They found that the population of giants with low [\al/Fe] has a distribution peaked at $\sim$2 Gyr and slopes gradually down to older ages, whereas the population with high [\al/Fe] peaks strongly at $\sim$11 Gyr. The ages of these populations were found to have limited overlap, with a transition at $\sim$8 Gyr, results which coincide with the observed ages in the solar neighborhood from \cite{fuhrmann1998,fuhrmann2011}. These results have been corraborated and expanded on by \cite{miglio2020}, who found ages for 5400 giants in the \Kepler field.
SA18 also found a surprising, small population of very young, \al-rich stars which, though possibly high-mass stars with genuinely young ages, could also indicate an unexpectedly large population of merger products---stars that have gained mass from a companion and therefore look younger than they actually are \citep{Martig2015,Chiappini2015}. Investigations of populations along other lines of sight have the potential to clarify the origin of this unusual cohort. 
The relationship shown by SA18 between age and \al-enrichment agrees very well with what might be expected if only considering the contribution of historic SNe rates to the interstellar medium. However, the sample of stars available from \cite{apokasc1, apokasc2}, and SA18's sub-sample, represent a population of stars very local to the Sun, only extending out to a distance of about 2 kpc. Therefore, whether these results are true for the \al-rich and \al-poor populations outside of the solar circle is still relatively uncertain.

In an effort to characterize the age distributions with asteroseismology outside of the solar vicinity, \cite{corogee} combined asteroseismology from two of the \textit{CoRoT} fields with APOGEE DR12 to get masses, radii, and ages using the Bayesian parameter estimation code PARAM \citep{param1,param2}. These fields, lying very close to the Galactic plane, provide a valuable sampling of the Milky Way's chemical history at a range of Galactic radii.

Building on work from \Kepler and \textit{CoRoT}, our paper takes advantage of the multiple Galactic sight-lines that the \ktwo mission \citep{k2mission} affords to better understand the age distributions of stars outside of the immediate solar neighborhood.
Following the failure of the \textit{Kepler} satellite's pointing, a new observing strategy was adopted to use the solar wind for partial stabilization. The re-purposed \textit{Kepler} mission, dubbed \textit{K2}, could no longer focus on the original \textit{Kepler} field. However, by virtue of its ecliptic orientation, it could now observe fields of view along the ecliptic for $\sim 80$ days at a time \citep{k2mission}. 
\ktwo therefore offers an expanded view of the Galaxy compared to the \textit{Kepler} prime mission, and thanks to a dedicated program to target red giants, the \ktwo Galactic Archaeology Program \citep[K2 GAP; ][]{stello+2015}, has yielded asteroseismic data beyond the solar vicinity \citep{stello+2017}.
As proof of the usefulness of \ktwo for Galactic archaeology, ages were found by \cite{rendle+2019} for stars in \ktwo campaigns 3 and 6. This includes a confirmation of the existence of young, \al-rich stars that otherwise are consistent with sharing the kinematic properties of old \al-rich stars.

In this paper, we look to further investigate the relationship between \al\ abundance and age by combining asteroseismic data obtained from observations of red giants observed by \ktwo with spectroscopic data from APOGEE DR16. Giants in the \ktwo fields, many of which were targeted at greater distances compared to the \Kepler giants, offer the opportunity to discover properties of \al-rich and \al-poor populations well beyond the solar circle. \S\ref{sec:data} describes how we selected the sample. In \S\ref{sec:method} we describe how we were able to obtain accurate age estimates for the sample of red giants. In \S\ref{sec:results} we discuss the ages that we found for the \al-rich and \al-poor popuations in the \ktwo fields and touch on some possible implications for the Galactic formation mechanisms mentioned above.

\section{Data Selection} \label{sec:data}

\begin{figure*}
	\plotone{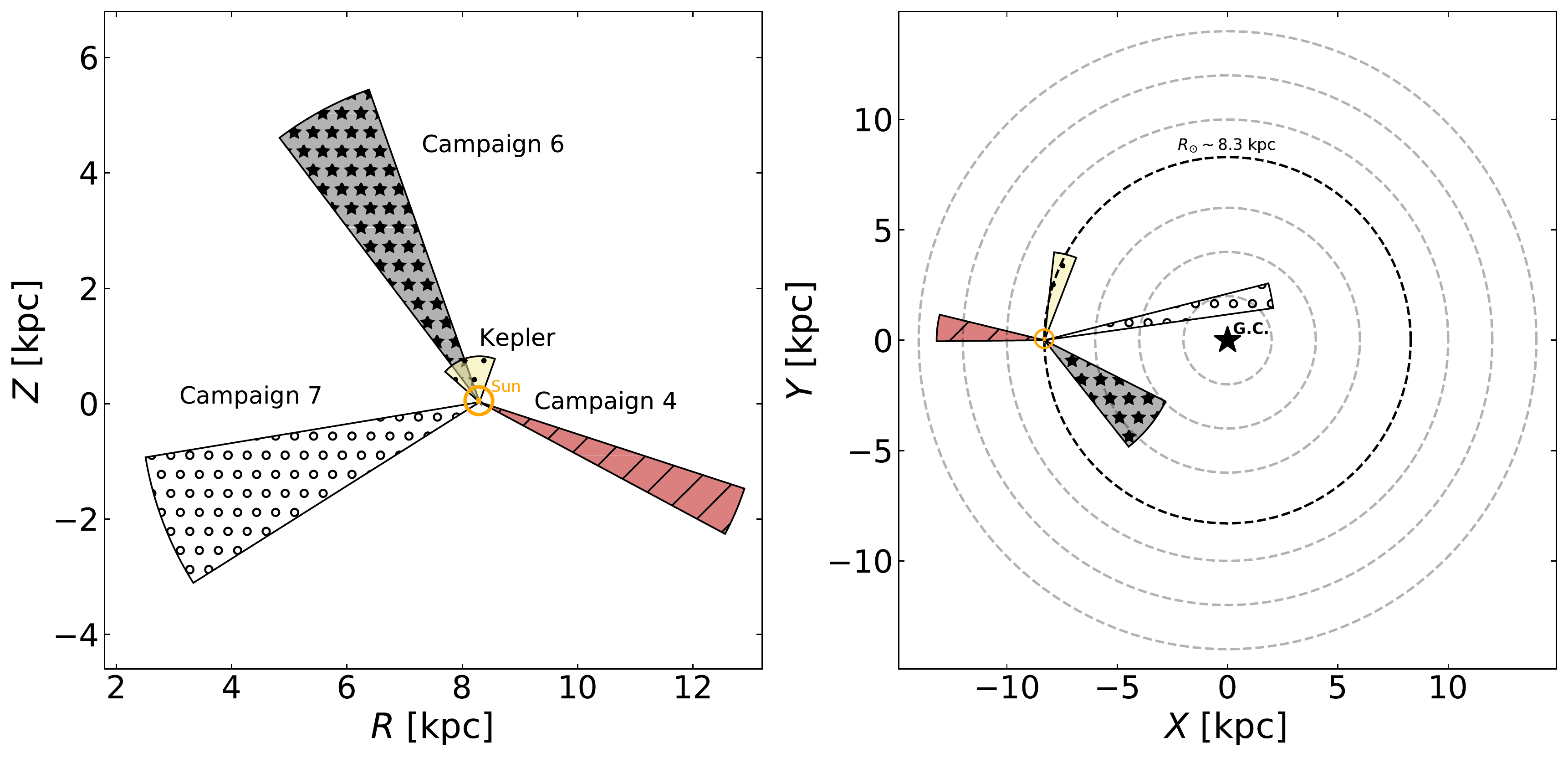}
	\caption{Schematics showing the approximate lines of sight for \ktwo campaigns 4, 6, and 7. The plot on the left shows the fields in the Galactocentric coordinates $Z$ (height above or below the Galactic plane) and $R$ (radial distance from the Galactic center). The plot on the right shows these fields in the Galactocentric $X$ and $Y$ coordinates. The Sun is approximated to be at $R=8.3$ kpc, $Z=0.027$ kpc, $X=-8.3$ kpc, and $Y=0$ kpc. Our sample contains 160 stars in C4, 411 stars in C6, and 233 stars in C7. \label{fig:schematic}}
\end{figure*}

\begin{figure}
	\plotone{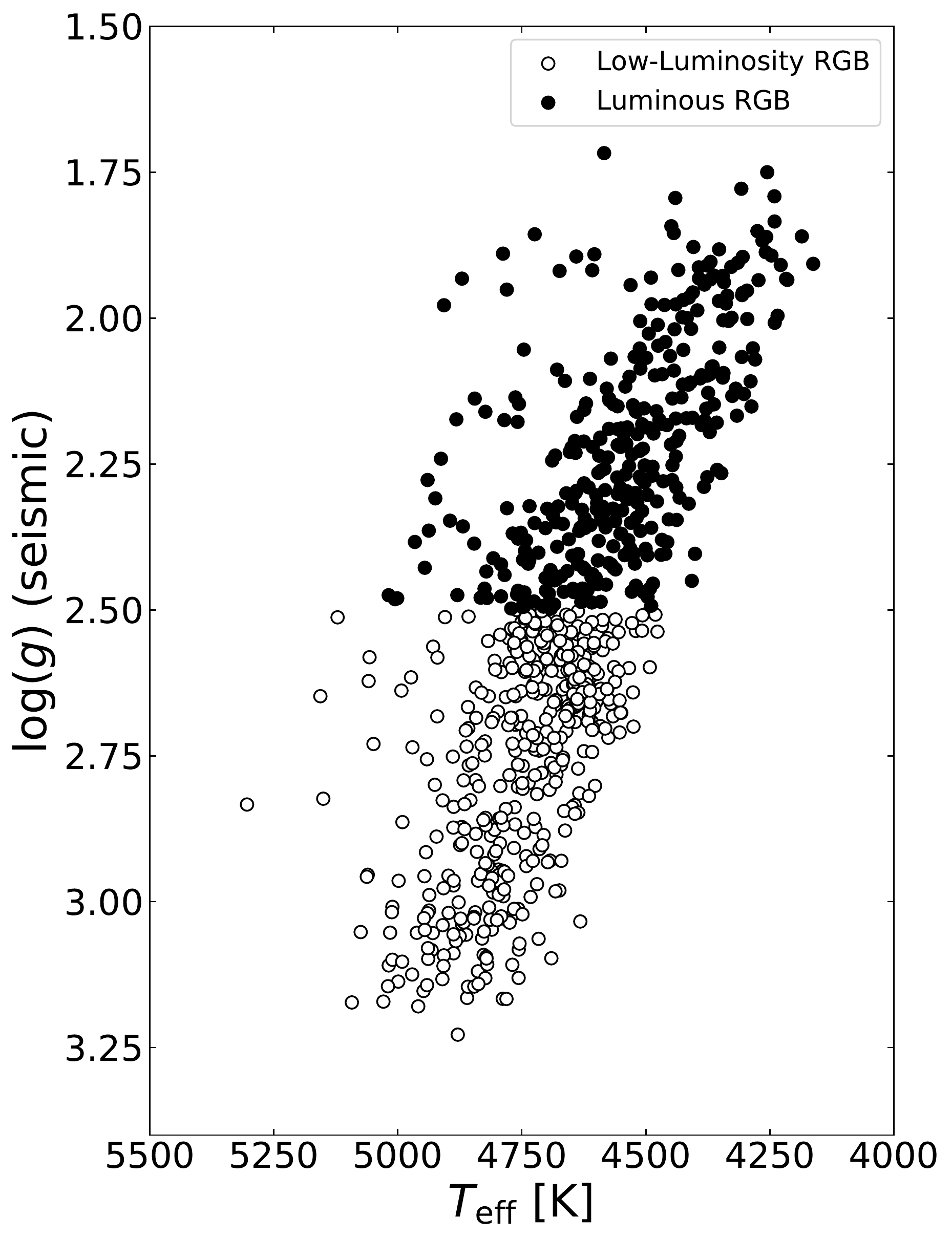}
	\caption{Kiel diagram for our sample of stars using effective temperatures from APOGEE and surface gravities calculated from the K2 GAP DR2 astereoseismic parameters. Low-luminosity giants are defined here as giants with $\log{(g)} > 2.5$. \label{fig:hrdiag}}
\end{figure}

The spectroscopic data for this work is from the 16th data release of the APOGEE survey \citep{majewski2017,apogeedr16}. This survey has been collecting high resolution IR spectra for hundreds of thousand of stars throughout the Milky Way. Access is provided to both the raw spectra and derived estimates for stellar parameters such as effective temperature ($T_{\mathrm{eff}}$), surface gravity ($g$), and chemical abundances via an automated pipeline. This includes estimates for metallicity ([Fe/H]) and \al-enrichment ([\al/M]\footnote{[$\alpha$/Fe] and [$\alpha$/M] are conceptually equivalent, with the [$\alpha$/M] parameter used by APOGEE measuring the ratio of $\alpha$ elements to the overall metallicity rather than just to iron.}).
The spectroscopic data in APOGEE DR16 was collected with the 2.5-meter Sloan Foundation Telescope \citep{gunn2006} at the Apache Point Observatory in New Mexico and the 2.5-meter du Pont Telescope \citep{bowen73} at Las Campanas Observatory in Chile as a part of SDSS-IV \citep{BlantonAndBershady2017}. APOGEE uses twin $R \sim 22,500$ H-band spectrographs \citep{wilson2019}.
APOGEE deliberately targeted red giants in the \Kepler and \ktwo fields to provide spectroscopic data for seismically detected stars. The selection of red giants in the \Kepler field is presented in \cite{zasowski2013,zasowski2017}, while the APOGEE program in the \ktwo fields is presented in \cite{zasowski2017} and Beaton et al. (in preparation). Briefly, the \ktwo program's stars were selected in a priority scheme as follows: (1) known planet hosts, (2) confirmed oscillators, (3) red giant targets in the GAP and not observed with HERMES, (4) red giants targeted in the GAP and observed with HERMES, (5) M-dwarf candidates, and (6) filler giants following the APOGEE-2 main red star sample \citep{zasowski2017}. Field centers were chosen to maximize the total priority and targets were selected in each field following the typical APOGEE fashion \citep{zasowski2013}.

The general scheme for extraction of the spectra, wavelength calibration, latfielding, and radial velocity measurements is described in \citet{Nidever2015}, with updates in \citet{holtzman2018} and \citet{jonsson2020}. Stellar parameters are obtained by processing the spectra through the APOGEE Stellar Parameters and Chemical Abundances Pipeline (ASPCAP) which infers these values by fitting the spectra to a grid of synthetic spectra \citep[for a description of the pipeline, see][]{aspcap}. \cite{jonsson2020} describes the updates to the pipeline for DR16, including the use of MARCS atmospheres for all stars with $T_{\rm eff} < 8000 \, \mathrm{K}$. Smith et al. (in preparation) describes the new line list used for construction of synthetic spectra in DR16.

Asteroseismic data for this work comes from the 2nd data release of \ktwo GAP \citep{k2gapdr2}. Over the mission lifespan, \ktwo made observations along 19 different lines-of-sight across the Galaxy which are referred to as campaigns 0-18. \ktwo GAP provides the analysis of asteroseismic targets in these fields. Because of the shorter time spent observing each of these fields versus the \textit{Kepler} field ($\sim80$ days versus $4$ years), the associated uncertainties with the asteroseismic parameters are significantly larger. However, the data from \ktwo offers a much more diverse positional sampling of stars in the Galaxy.
The full \ktwo GAP project covers \ktwo campaigns 1-18 over a series of three data releases. Data Release 1 \citep{stello+2017} provided analysis for $\sim 1200$ stars in \ktwo campaign 1 and worked as a proof of concept for the project. Data Release 2, used in this work, provides analysis of $\sim 4500$ stars in \ktwo campaigns 4, 6, and 7. These three campaigns were chosen as they probe three lines of sight that are both distinct from each other and from the \Kepler field (Figure \ref{fig:schematic}). Data Release 3 will provide the results for the remaining \ktwo campaigns.

Following methods developed for \textit{Kepler} \citep{apokasc2}, \ktwo GAP DR2 improves on \ktwo GAP DR1 by providing asteroseismic data that have been averaged across results from up to six independent pipelines. These pipelines are A2Z+ \citep[based on][]{A2Z}, BHM \citep{ElsworthBHM,BHM}, CAN \citep{CAN}, COR \citep{COR}, SYD \citep{SYD}, and BAM \citep{BAM}. In this paper, we only consider giants in this catalog for which at least two of these pipelines returned values for both $\nu_{\mathrm{max}}$ and $\Delta\nu$, and take the mean of these results. We include these mean values along with all of the associated pipeline values in our catalog.

To investigate trends in the age-abundance relationship with Galactic position, we adopted Bayesian distance estimates based on \textit{Gaia} Data Release 2 \citep{gaia1,gaia2} parallaxes from \citet{BailerJones2015}.
Each star's height above the Galactic plane ($Z$), radial distance from the center of the Galaxy ($R$), and Galactocentric $X$ and $Y$ coordinates were computed using \texttt{Astropy}\footnote{https://www.astropy.org} \citep{astropy:2013, astropy:2018}.

We have extensive asteroseismic data sets for both shell H-burning red giant branch (RGB) stars and core He-burning red clump (RC) stars. We restrict ourselves to RGB stars because the mapping between mass and age for RC stars is complicated by the known existence of mass-loss between the RGB and RC phases, so that the present-day RC asteroseismic mass is systematically biased compared to its birth mass \citep[e.g.,][]{casagrande+2016}.

The state-of-the-art for RGB vs. RC classifications is to determine a star's evolutionary state through asteroseismology \citep[e.g.,][]{bedding+2011}. However, this method requires very long time domain data. Though these classifications are difficult to do with the short \ktwo time series, they have been successfully made using the \Kepler data by \cite{apokascstates}.

To distinguish between RGB and RC stars in \textit{K2}, we used a spectroscopic evolutionary state classification similar to that of \cite{Bovy2014}. A temperature-, surface gravity-, and abundance-dependent cut to separate RGB and RC stars was found using the evolutionary states from \cite{apokascstates}. First, we fit for $\alpha$, $\beta$, and $\gamma$ that define a "reference" temperature given by
\begin{equation} \label{eq:tref}
    T_{\mathrm{ref}} = \alpha + \beta\, \mathrm{[Fe/H]_{\mathrm{RAW}}} + \gamma\, (\log{(g)}_{\mathrm{SPEC}} - 2.5)
\end{equation}
through a non-linear least squares fit for stars classified as RGB according to APOKASC-2 asteroseismology. [Fe/H]$_{\mathrm{RAW}}$ and $\log{(g)}_{\mathrm{SPEC}}$ are the uncorrected values for metallicity and surface gravity, respectively, given in the APOGEE DR16 catalogue. The fitting results were found to have approximate values of $\alpha = 4383.148$ K, $\beta = -235.136$ K/dex, and $\gamma = 532.659$ K/dex. A line was then fit through the approximate ridgeline in [C/N]$_{\mathrm{RAW}}$ vs. $T_{\mathrm{eff}}^{\mathrm{SPEC}} - T_{\mathrm{ref}}$ space for which 98\% of stars to the right of the line were classified as RGB. This ensures that the contamination from RC stars in our spectroscopic classifications is negligible. The finalized classification criteria in Table \ref{tab:states} were then used to pick out the stars in our \ktwo sample that are most likely to be on the RGB. Hereafter in this work, all references to "giants" refer specifically to RGB stars unless otherwise stated.
\begin{table*}
\caption{Grid for if a star is classified as a red giant for different ranges of $\log{(g)}_{\mathrm{SPEC}}$ and [C/N]$_{\mathrm{RAW}}$. Stars are classified as red giants if the statement for the given range of conditions is true for that star. $\Delta T = T_{\mathrm{eff}}^{\mathrm{SPEC}} - T_{\mathrm{ref}}.$ \label{tab:states}}
\begin{tabular}{|c||c|c|}
 \hline
 & $3.5 > \log{(g)}_{\mathrm{SPEC}} > 2.38$ & $\log{(g)}_{\mathrm{SPEC}} < 2.38$ \\
 \hline\hline
 $\mathrm{[C/N]}_{\mathrm{RAW}} > -0.2969$ & $\mathrm{[C/N]}_{\mathrm{RAW}} < 0.0209 - 0.5352 \mathrm{[Fe/H]}_{\mathrm{RAW}} - 0.0029 \Delta T$ & True \\
 \hline
 $\mathrm{[C/N]}_{\mathrm{RAW}} < -0.2969$ & $150 > 182.662 \mathrm{[Fe/H]}_{\mathrm{RAW}} + \Delta T$ & True \\
 \hline
\end{tabular}
\end{table*}

From the parameters provided by \ktwo GAP and APOGEE, we were able to calculate values for asteroseismic surface gravity ($\log{(g)}_{\mathrm{seis}}$), for mass, and for radius. $\log{(g)}_{\mathrm{seis}}$ was calculated using $\nu_{\text{max}}$ and $T_{\text{eff}}$ in the scaling relation \citep{brown1991,kjeldsen&bedding1995}:
\begin{equation} \label{eq:numaxscale}
    \left (\frac{\nu_{\text{max}}}{\nu_{\rm{max,}\odot}}\right) \approx \left( \frac{g}{g_{\odot}}\right) \left(\frac{T_{\text{eff}}}{T_{\text{eff},\odot}}\right)^{-1/2},
\end{equation}
with solar reference values of $\nu_{\text{max},\odot}$ = 3076 $\mu$Hz, $T_{\text{eff},\odot}$ = 5772 K, and $g_{\odot}$ = 27400 cm/s$^2.$\footnote{This value agrees to within $0.0003$ dex with the recent standard solar values adopted in \citealt{mamajek+2015}.} Values for mass and radius can be found by combining Equation~\ref{eq:numaxscale} with the scaling relation for $\Delta\nu$ \citep{ulrich1986,kjeldsen&bedding1995},
\begin{equation} \label{eq:deltanuscale}
    \Delta\nu \propto M^{1/2} R^{-3/2}.
\end{equation}
Doing this gives that
\begin{equation} \label{eq:seismass}
    \left(\frac{M}{M_{\odot}}\right) \approx \left (\frac{\nu_{\text{max}}}{f_{\nu_{\text{max}}}\nu_{\rm{max,}\odot}}\right)^3 \left( \frac{T_{\text{eff}}}{T_{\text{eff},\odot}}\right)^{3/2} \left( \frac{\Delta\nu}{f_{\Delta \nu}\Delta \nu_{\odot}}\right)^{-4}
\end{equation}
and
\begin{equation} \label{eq:seisradius}
    \left( \frac{R}{R_{\odot}}\right) \approx \left (\frac{\nu_{\text{max}}}{f_{\nu_{\text{max}}}\nu_{\rm{max,}\odot}}\right) \left( \frac{T_{\text{eff}}}{T_{\text{eff},\odot}}\right)^{1/2} \left(\frac{\Delta\nu}{f_{\Delta\nu} \Delta\nu_{\odot}}\right)^{-2}.
\end{equation}
We adopt $\Delta \nu_{\odot} =$ 135.146 $\mu$Hz. We also compute theoretically-motivated corrections to the observed $\Delta \nu$ values, $f_{\Delta\nu}$, according to \cite{sharma+2016}. There is increasing empirical evidence that these and similar corrections from the literature \citep[e.g.,][]{white+2011,guggenberg+2017} result in better agreement with independent estimates of stellar parameters \citep[e.g.,][]{huber+2017}. Evaluating $f_{\Delta\nu}$ requires mass, radius, temperature, evolutionary state, and metallicity. The bulk metallicities are adjusted for this purpose to account for non-solar alpha abundances according to the \cite{salaris+1993} prescription: 
$\mathrm{[Fe/H]}' = \mathrm{[Fe/H]} + \log_{10}(0.638 \times 10^{\mathrm{[\alpha/M]}} + 0.362)$. 
We assume that there are no corrections needed to the observed $\nu_{\rm{max}}$ values (i.e., $f_{\nu_{\text{max}}} = 1$), pending further empirical constraints and theoretical understanding of the $\nu_{\rm{max}}$ scaling relation (see discussion in, e.g., \citealt{apokasc2}).

The sample was then further limited to stars with [$\alpha$/M] values between 0.0 and 0.4 dex, [Fe/H] values between -2.0 and 0.6 dex, and masses between 0.6 and 2.6 $M_{\odot}$ in order to ensure meaningful parameter values that fit within the parameter space defined by the tracks that we used for estimating ages (\S\ref{sec:method}). It should be noted that it is still unclear how well the asteroseismic scaling relations work for low-metallicity stars with $\mathrm{[Fe/H]} < -1$ \citep[e.g.,][]{Epstein2014,Valentini2019}, however, the inclusion of these few objects in our sample does not have a noticeable impact on our results for the bulk populations. As a final quality check, we consider low- and high- luminosity RGB stars separately in \S\ref{sec:results}. High-luminosity giants appear to suffer from measurement systematics \citep{Mosser2013,apokasc2,joelradius} and so, out of an abundance of caution, we employ a conservative surface gravity cut to separate our sample: stars with  $\log{(g)}_{\mathrm{seis}} < 2.5$ are classified as luminous giants and stars with $\log{(g)}_{\mathrm{seis}} > 2.5$ are classified as low-luminosity giants (Figure \ref{fig:hrdiag}).

\begin{figure*}
	\plotone{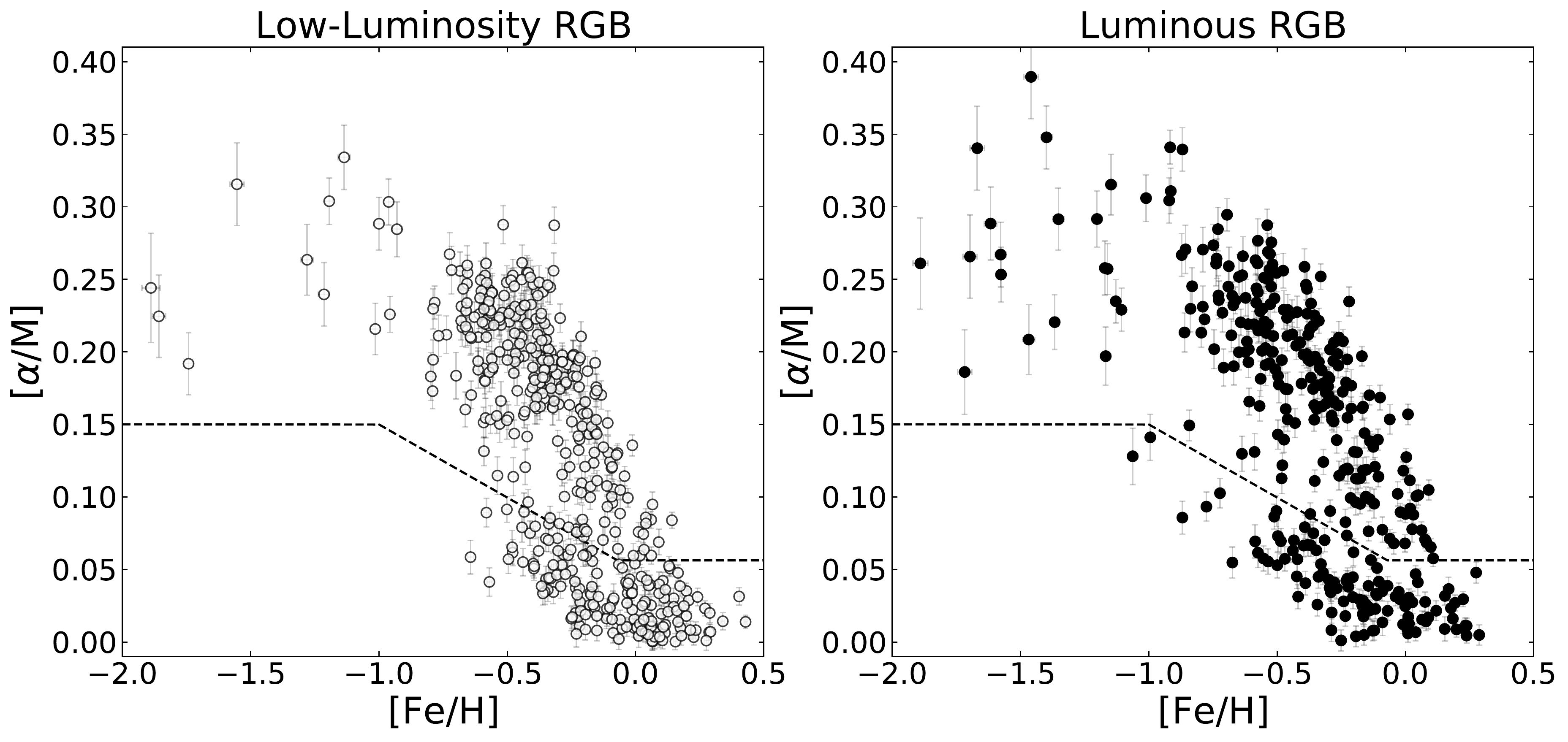}
	\caption{[\al/M] vs. [Fe/H] for the low-luminosity and luminous giants in our sample. Our sample is further split into the categories of \al-rich and \al-poor, which is defined by the dashed line in the plots. This division was defined by-eye based on the ridge-line between the groups of points in the data. This cut is similar to that made by e.g. \cite{SilvaAguirre2018} and \cite{Weinberg2019}. This line is defined as ${\rm [\alpha/M]}=0.15$ for ${\rm [Fe/H]}<-1.0$, ${\rm [\alpha/M]}=0.056$ for ${\rm [Fe/H]}>-0.07$, and ${\rm [\alpha/M]}=-0.1 \cdot {\rm [Fe/H]}+0.049$ for $-1.0 \leq {\rm [Fe/H]} \leq -0.07$. \label{fig:avm1}}
\end{figure*}

Lastly, the sample was divided into the categories of \al-rich and \al-poor by approximately drawing a line through the ridge-line between the two populations, as seen in Figure~\ref{fig:avm1}.

\section{Age Determination} \label{sec:method}

Obtaining reliable ages for RGB stars is primarily dependent on having accurate estimates for stellar mass and secondarily on harmoniously-calibrated values for chemical compositions.
Because the lifetime in the RGB phase is short, the surface gravity has only a minor impact on the derived age, especially on the upper RGB.
Because of mechanical challenges in incorporating the age dependence tied to surface gravity, which is multi-valued for the RGB, we simply bracket our sample into two surface gravity bins and take the age range within these bins as a (very small) additional source of uncertainty.

At fixed surface gravity and composition, age is strongly sensitive to effective temperature.
In grid modeling, the derived age combines this information with asteroseismic properties; for an example see SA18. However, there are large random and systematic uncertainties in this age estimate due to effective temperature offsets. On the giant branch locus, an error budget of $50 {\rm\, K}$ in temperature would produce a random age uncertainty of $\sim70\%$.
A mismatch between the true locus of the seismic data and stellar models can also produce very large zero-point offsets and composition-dependent systematic shifts \citep{Tayar2017}. We therefore adopt the methodology of \citet{apokasc2}, and do not directly incorporate classical age constraints from HR diagram position.
We note that this choice also makes the ages more replicable, and that others can use different choices of stellar models with these data to infer ages with a grid modeling approach if they so choose.

We used stellar evolutionary tracks, generated with the Yale Rotating Evolution Code, from \cite{yrec} with updates from \cite{tracks}. From these tracks we created three sets of grids for $\log{(g)}$ of 3.30, 2.50, and 1.74, with columns for $\log{(\mathrm{Mass})}$, [Fe/H], [$\alpha$/Fe], and $\log{(\mathrm{age})}$. These values for $\log{(g)}$ were chosen to approximately bracket the low-luminosity giant branch (3.30 and 2.50) and the upper giant branch (2.50 and 1.74). We made these grids regular along the $\log{(\mathrm{Mass})}$, [Fe/H], and [$\alpha$/Fe] axes by linearly interpolating to ages at locations where there were gaps in the tracks. Given a star's values and associated uncertainties for each of these parameters, a Monte Carlo method was used to calculate 500 age estimates for each star through 4-point Lagrange interpolation within the grid. The reported results for a given star is the median of these values along with a lower and an upper 1-$\sigma$ uncertainty reflecting the 16th and 84th percentiles of the Monte Carlo results, respectively.

\begin{figure}
    	\plotone{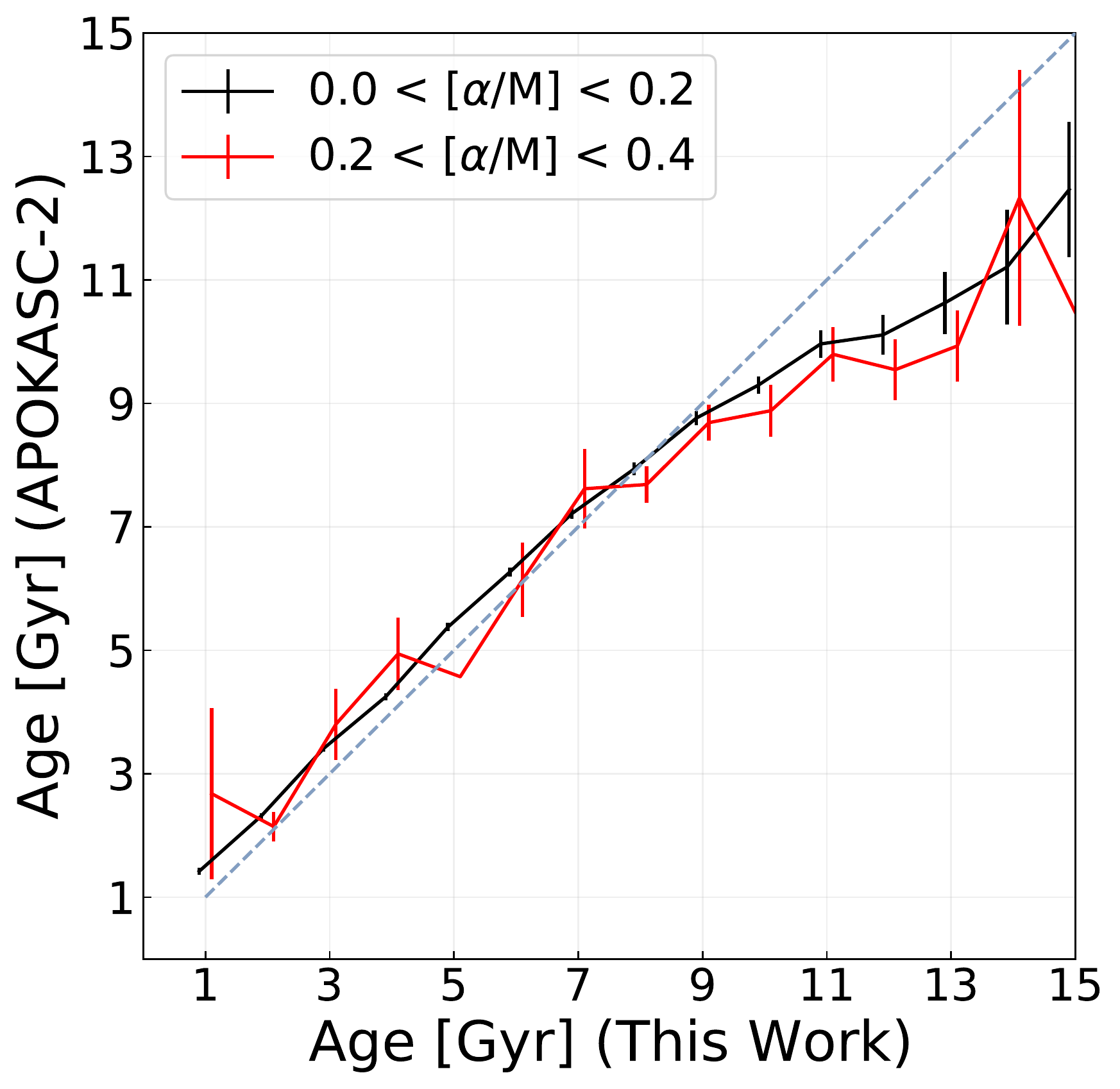}
	\caption{A comparison of the ages reported in the APOKASC-2 catalog vs. the ages calculated for this work for 2407 RGB stars from \cite{apokasc2}. Error bars represent the standard deviation of the mean for bins that are 1 Gyr wide on the $x$-axis. \label{fig:fig4}}
\end{figure}
Figure \ref{fig:fig4} shows a comparison between the ages calculated for 2407 RGB stars in the \Kepler field using this method and the ages reported for the same giants in the APOKASC-2 catalog \citep{apokasc2}. 
There are a few main systematic effects, aside from differences in technique, that lead to the slight disagreement in the ages and the age zero-point.
First, though \cite{apokasc2} used a similar method to estimate the ages of these stars, the stellar tracks and isochrones used did not take $\alpha$ enhancement into account when generated. Rather, the lookup metallicity was corrected by adjusting [Fe/H] by $+0.625$ [$\alpha$/M].
At a fixed [Fe/H], an increase in [\al/M] corresponds to an increase in a star's opacity. Increasing a star's opacity means also increasing its radius, therefore lowering the star's core temperature and extending the amount of time that it takes to burn through core hydrogen on the main sequence. 
Therefore, not considering [\al/M] in stellar models has the effect of underestimating the ages of stars with [\al/M] $>0$, with the biggest effect being for the ages of low-mass stars. The correction on metallicity used by \cite{apokasc2} is not able to fully account for this effect. On average, this affects the ages of stars younger than about 8 Gyr by approximately $+1\%$ and above 8 Gyr by $+8\%$. The second systematic is that, in order to make the ages in the \Kepler field more concordant with those from the \ktwo fields, we recalculated $f_{\Delta\nu}$ for this sample in the same manner that it was calculated for \ktwo GAP. This adjustment leads to changes in mass by an average of about $+2\%$ to $+3\%$, which corresponds to changes of $-6\%$ to $-9\%$ in age.
In order to make masses agree with those from open clusters, \cite{apokasc2} applied a uniform shift in $f_{\nu_{\text{max}}}$ (and therefore the sample's asteroseismic masses). Correcting the zero-point offset from our reformulation of values for $f_{\Delta\nu}$ so that the masses are on the same open cluster scale, therefore, required us to scale masses for the \textit{Kepler} sample down by a factor of $1.018$, which translates to scaling ages up by a factor of about $1.018^3 = 1.055$ (or about 6\%).
In addition to these effects, there are also going to be changes associated with the difference between the abundance estimates in APOGEE DR14 (which was used for the ages by \citealt{apokasc2}) and APOGEE DR16 \citep[changes between these data releases are discussed in][]{jonsson2020}.

Another check we can perform on the data is to see if our age distributions for \Kepler stars reproduce the features seen by SA18. The ages published by SA18 are computed using the Bayesian stellar parameter estimation package \texttt{BASTA} \citep{silva_aguirre+2015,silva_aguirre+2017}. The \Kepler age distribution globally reproduces the features of that from SA18, including the small young, \al-rich population. This indicates that adopting a grid modeling approach does not induce a large change in derived ages.
We also note that our \Kepler age distribution is consistent with recent estimates of the thick disk age using detailed asteroseismic modeling of \Kepler data \citep{montalban+2020}.

\section{Results and Discussion} \label{sec:results}

\begin{figure*}
    \plotone{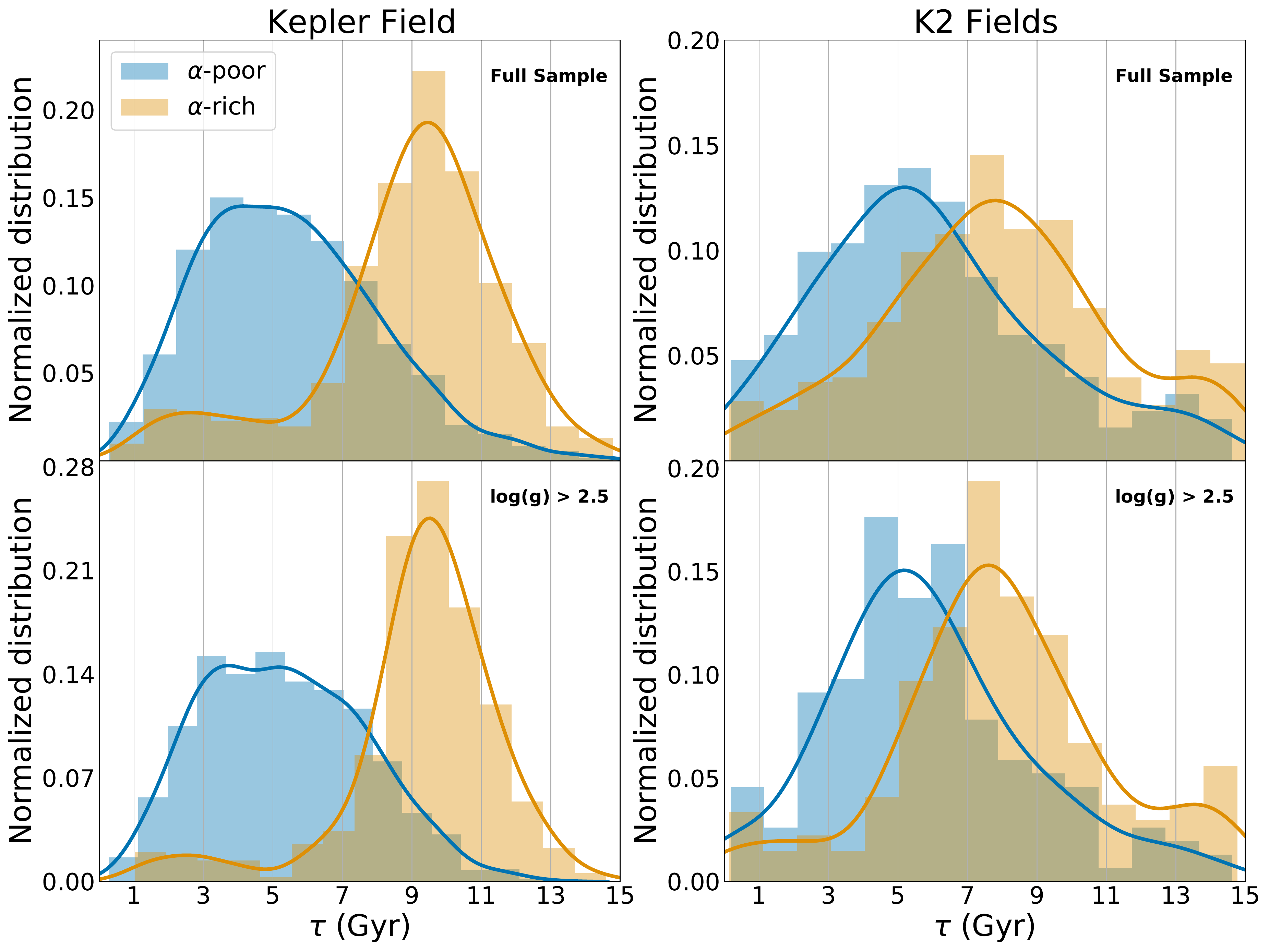}
    \caption{Distributions of the age estimates for the Kepler field (left) and \ktwo fields (right) for both the $\alpha$-poor (blue) and $\alpha$-rich (orange) populations. The top row shows the distributions for the full sample for the respective fields and the bottom row shows the distributions for the low-luminosity sample ($\log{(g)}_{\mathrm{seis}} > 2.5$). In the \Kepler field, there are 691 (389) \al-rich and 1714 (1228) \al-poor stars in the full (low-luminosity) sample. In the \ktwo fields, there are 534 (291) \al-rich and 270 (160) \al-poor stars in the full (low-luminosity) sample.
    \label{fig:kepk2}}
\end{figure*}
\begin{figure*}
    \plotone{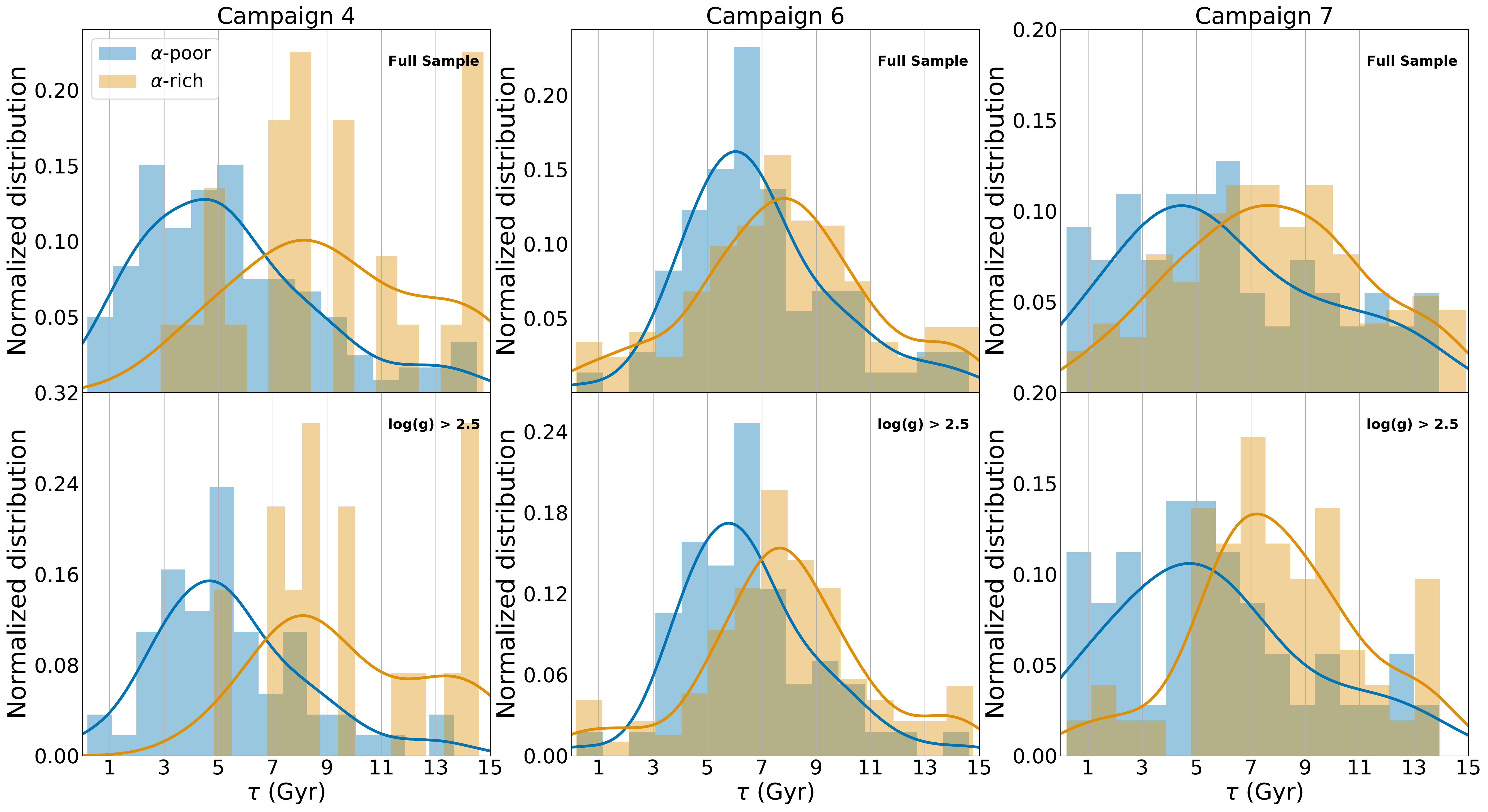}
    \caption{Distributions of the age estimates for both the $\alpha$-poor (blue) and $\alpha$-rich (orange) populations for each \ktwo field individually. The top row shows the distributions for the full sample for the respective fields and the bottom row shows the distributions for the low-luminosity sample ($\log{(g)}_{\mathrm{seis}} > 2.5$).
    In C4, there are 32 (22) \al-rich and 128 (62) \al-poor stars in the full (low-luminosity) sample. In C6, there are 335 (208) \al-rich and 76 (59) \al-poor stars in the full (low-luminosity) sample. In C7, there are 167 (61) \al-rich and 66 (39) \al-poor stars in the full (low-luminosity) sample.
    \label{fig:k2all}}
\end{figure*}
The panels in Figure \ref{fig:kepk2} show the age results from both the \Kepler and \ktwo fields. Results are shown for both the full samples and for the low-luminosity giants in each field. Figure \ref{fig:k2all} shows these same results for each \ktwo campaign individually. Gaussian kernel density functions were drawn over each distribution using the \texttt{kdeplot} function from the \texttt{seaborn}\footnote{https://seaborn.pydata.org} Python package \citep{seaborn}. We can see that, though the inclusion of the luminous giants does not have a significant impact on the locations of the peaks for the underlying kernel density estimates they do contribute a noticeable degree of scatter to the results..

\begin{deluxetable*}{|c|c|c|c|c|}
  \tablecaption{This table gives information for the kernel density estimation (kde) and the underlying data for the age-distributions of the low-luminosity $\alpha$-rich populations in each field (shown in Figures \ref{fig:kepk2} and \ref{fig:k2all}). The value for the \al-rich age width is calculated as $\mathrm{FWHM}/2.355$, where $\mathrm{FWHM}$ is the full width at half maximum around the peak in the kde. The mean error is the average error in the age estimate for only the $\alpha$-rich stars in the indicated field. The possible systematic uncertainties come from the interpolation error (-1\%) and from the systematic uncertainty due to the \ktwo $\nu_{\rm max}$ scale (+15\%).
\label{tab:histtable}}
  \tablehead{Field & $\alpha$-rich Peak Age (Gyr) & $\alpha$-rich Peak Age Width (Gyr) & Average Age Uncertainty (Gyr) & Possible Sys. Uncertainties (\%)}
\startdata
 Kepler & 9.53 & 1.40 & 1.46 & -1 \\
 \hline
 K2 C4 & 8.08 & 3.83 & 2.92 & +14\\
 \hline
 K2 C6 & 7.62 & 2.16 & 2.23 & +14\\
 \hline
 K2 C7 & 7.24 & 2.63 & 2.40 & +14\\
 \hline
 all K2 & 7.61 & 2.17 & 2.32 & +14\\
\enddata
\end{deluxetable*}
\begin{splitdeluxetable*}{cclccccclBcccccccccccBccccccccccc}
  \tablecaption{The partial data table for \ktwo campaigns 4, 6, and 7, including our ages. The complete table is available in machine-readable format in the online journal, which also includes the asteroseismic values from each of the individual pipelines. \label{tab:k2_table}}
  \tabletypesize{\footnotesize}
  \tablehead{\colhead{K2 EPIC ID} &
\colhead{K2 Campaign} &
\colhead{APOGEE ID} &
\colhead{$\tau$ (Gyr)} &
\colhead{$\sigma^{-}(\tau)$} &
\colhead{$\sigma^{+}(\tau)$} &
\colhead{RA (deg.)} &
\colhead{Dec (deg.)} &
\colhead{{\it Gaia} Designation} &
\colhead{$T_{\rm eff}$ (K)} &
\colhead{$\sigma(T_{\rm eff})$} &
\colhead{$\log{(g)}$ (cgs)} &
\colhead{$\sigma(\log{(g)})$} &
\colhead{${\rm [\alpha/M]}$} &
\colhead{$\sigma({\rm [\alpha/M]})$} &
\colhead{${\rm [Fe/H]}$} &
\colhead{$\sigma({\rm [Fe/H]})$} &
\colhead{$\bar{\nu}_{\rm max}$ (${\rm \mu Hz}$)} &
\colhead{$\sigma(\bar{\nu}_{\rm max})$} &
\colhead{$\Delta \bar{\nu}$ (${\rm \mu Hz}$)} &
\colhead{$\sigma(\Delta \bar{\nu})$} &
\colhead{$\log{(g)}_{\rm seis}$ (cgs)} &
\colhead{$\sigma(\log{(g)}_{\rm seis})$} &
\colhead{$M$ ($M_\odot$)} &
\colhead{$\sigma(M)$} &
\colhead{$d$ (kpc)} &
\colhead{$\sigma(d)$} &
\colhead{$X$ (kpc)} &
\colhead{$Y$ (kpc)} &
\colhead{$Z$ (kpc)} &
\colhead{$R$ (kpc)}}
\startdata 
210495151.0 &            4 &  2M03593912+1519463 &  9.94 &     1.81 &     2.28 &  59.913042 &  15.329524 &    Gaia DR2 40120290241918848 &      4797.24 &         106.18 &         2.87 &           0.07 &            0.23 &              0.01 &        -0.58 &           0.01 &            108.47 &           0.82 &           10.36 &         0.10 &       2.94 &         0.01 &    0.96 &      0.05 &          1.11 &            0.05 &      -9.28 &       0.07 &      -0.48 &       9.28 \\
 210610252.0 &            4 &  2M04211626+1706358 &  9.48 &     1.71 &     2.48 &  65.317758 &  17.109962 &  Gaia DR2 3313915749626906752 &      4750.61 &          85.03 &         2.61 &           0.06 &            0.04 &              0.01 &        -0.23 &           0.01 &             44.15 &           0.60 &            5.19 &         0.04 &       2.55 &         0.01 &    1.01 &      0.06 &          1.24 &            0.08 &      -9.44 &       0.04 &      -0.45 &       9.44 \\
 210483090.0 &            4 &  2M03585776+1507226 &  2.98 &     0.36 &     0.38 &  59.740697 &  15.122972 &    Gaia DR2 40110463356775296 &      4958.78 &          89.15 &         3.15 &           0.06 &            0.04 &              0.01 &        -0.01 &           0.01 &            183.08 &           1.38 &           14.02 &         0.02 &       3.18 &         0.01 &    1.45 &      0.05 &          2.35 &            2.66 &     -10.37 &       0.15 &      -1.07 &      10.38 \\
 210505442.0 &            4 &  2M04003325+1530162 &  4.41 &     1.17 &     1.65 &  60.138593 &  15.504482 &    Gaia DR2 40147159557282560 &      4889.04 &          91.34 &         2.84 &           0.07 &            0.06 &              0.01 &        -0.48 &           0.01 &             68.83 &           0.68 &            7.10 &         0.14 &       2.75 &         0.01 &    1.19 &      0.11 &          1.14 &            0.05 &      -9.31 &       0.07 &      -0.49 &       9.31 \\
 210720697.0 &            4 &  2M04220023+1839546 &  3.62 &     0.66 &     0.88 &  65.500938 &  18.665140 &    Gaia DR2 47563498628339072 &      4726.70 &          83.11 &         2.58 &           0.06 &            0.01 &              0.01 &        -0.09 &           0.01 &             52.67 &           0.59 &            5.50 &         0.05 &       2.63 &         0.01 &    1.35 &      0.08 &          1.58 &            0.12 &      -9.77 &       0.08 &      -0.55 &       9.77 \\
\enddata
\end{splitdeluxetable*}
\begin{splitdeluxetable*}{clccccccccccBcccccccccccc}
    \tablecaption{The partial data table for RGB stars in the \Kepler field. The complete table is available in machine-readable format in the online journal. \label{tab:kep_table}}
\tabletypesize{\scriptsize}
\tablehead{
\colhead{KIC} &
\colhead{APOGEE ID} &
\colhead{$\tau$ (Gyr)} &
\colhead{$\sigma^{-}(\tau)$} &
\colhead{$\sigma^{+}(\tau)$} &
\colhead{APOKASC2 $\tau$ (Gyr)} &
\colhead{$T_{\rm eff}$ (K)} &
\colhead{$\sigma(T_{\rm eff})$} &
\colhead{$\log{(g)}$ (cgs)} &
\colhead{$\sigma(\log{(g)})$} &
\colhead{${\rm [\alpha/M]}$} &
\colhead{$\sigma({\rm [\alpha/M]})$} &
\colhead{${\rm [Fe/H]}$} &
\colhead{$\sigma({\rm [Fe/H]})$} &
\colhead{$\nu_{\rm max}$ (${\rm \mu Hz}$)} &
\colhead{$\sigma(\nu_{\rm max})$} &
\colhead{$\Delta \nu$ (${\rm \mu Hz}$)} &
\colhead{$\sigma(\Delta \nu)$} &
\colhead{$M$ ($M_\odot$)} &
\colhead{$\sigma(M)$} &
\colhead{$X$ (kpc)} &
\colhead{$Y$ (kpc)} &
\colhead{$Z$ (kpc)} &
\colhead{$R$ (kpc)}
}
\startdata
 1027337 &  2M19252021+3647118 &   5.49 &     1.09 &     1.27 &          6.28 &      4635.50 &          78.02 &         2.76 &           0.05 &            0.01 &              0.01 &         0.20 &           0.01 &            73.97 &               0.67 &          7.09 &            0.09 &        1.29 &          0.08 &      -7.81 &       1.31 &       0.26 &       7.92 \\
 1296068 &  2M19264481+3658152 &  10.31 &     1.34 &     1.68 &         10.42 &      4586.52 &          93.37 &         2.61 &           0.05 &            0.05 &              0.01 &        -0.02 &           0.01 &            59.38 &               0.53 &          6.32 &            0.02 &        1.04 &          0.04 &      -7.72 &       1.57 &       0.31 &       7.88 \\
 1429505 &  2M19225688+3702125 &   6.42 &     0.96 &     1.20 &          7.36 &      4682.15 &          87.07 &         2.64 &           0.06 &            0.02 &              0.01 &        -0.11 &           0.01 &            55.41 &               0.61 &          5.91 &            0.02 &        1.15 &          0.05 &      -7.88 &       1.13 &       0.24 &       7.96 \\
 1431059 &  2M19243068+3701290 &   7.71 &     1.22 &     1.68 &          6.98 &      4802.11 &         106.14 &         3.07 &           0.06 &            0.00 &              0.01 &         0.06 &           0.01 &           170.84 &               1.54 &         13.86 &            0.08 &        1.14 &          0.05 &      -7.83 &       1.26 &       0.26 &       7.93 \\
 1433730 &  2M19265020+3703054 &   1.35 &     0.16 &     0.21 &          1.43 &      4732.06 &          83.20 &         2.53 &           0.06 &            0.00 &              0.01 &        -0.08 &           0.01 &            40.01 &               0.36 &          4.17 &            0.02 &        1.79 &          0.07 &      -7.91 &       1.06 &       0.22 &       7.98 \\
\enddata
\end{splitdeluxetable*}
Table \ref{tab:histtable} summarizes these results for the $\alpha$-rich giants in each field. Our full \ktwo data sample, including age estimates for individual stars, are provided in Table \ref{tab:k2_table}.
Table \ref{tab:kep_table} contains our ages for stars in the \Kepler field.

Comparing our results for these \ktwo fields with the results from the Kepler field brings to light two interesting differences. First, though the median age of the \al-rich population is strongly peaked at a single age in both of these samples, the populations in the \ktwo fields are found to be at an age about 2 Gyr younger than what was found in the \Kepler field. Secondly, there seems to be much less of a difference between the ages of the \al-rich and \al-poor populations in the \ktwo field.

Of course, it is still possible that there are systematics at play partially driving these discrepancies. The \ktwo asteroseismic parameters used in this work yield radii that may be up to $\sim 2\%$ larger than radii computed based on \textit{Gaia} DR2 parallaxes \citep{k2gapdr2}. Assuming this offset is due to slightly different $\nu_{\rm max}$ scale compared to APOKASC-2, this implies that the \ktwo masses may be too large by $\sim 5\%$, which would allow for the \ktwo ages to be $\sim 15\%$ younger than the \Kepler ages. In this sense, the age gap we find between \ktwo and \Kepler may be related to the fundamental scale of the \ktwo asteroseismic data. At this time, we are limited in our ability to calibrate the \ktwo asteroseismic scale with \textit{Gaia} data due to the \textit{Gaia} DR2 parallax zero-point varying across the sky. The next data release of \textit{Gaia} data should be less affected by the zero-point, and will be very useful in solidifying the \ktwo asteroseismic scale \citep[see also][]{Khan2019}.

There is, however, some reason to believe that an intermediately aged \al-rich population could be physical: \cite{lian+2020} discuss a population of young, \al-rich stars in the outer disk, suggesting that there should have also been mechanisms in place to make the intermediate age populations observed in \textit{K2}.
In addition, the result from \cite{corogee} of a young $\alpha$-rich population with similar ages to the corresponding $\alpha$-poor population in the \textit{CoRoT} fields seems to also indicate that an intermediate aged \al-rich population is not unthinkable. At the moment, the comparison of our results to \citealt{corogee}'s can only be qualitative. As opposed to SA18, \cite{apokasc2}, \cite{k2gapdr2}, and this work, \cite{corogee} does not apply theoretically-motivated $\Delta\nu$ corrections to their asteroseismic results, which could results in an offset of around 30\% in age.

It should be noted that, though there is significantly more spread in the ages around the \al-rich peak in the \ktwo fields, there is no evidence that this spread is due to anything other than the larger uncertainties associated with this data.

\begin{figure}
    \begin{center}\includegraphics[height=5in]{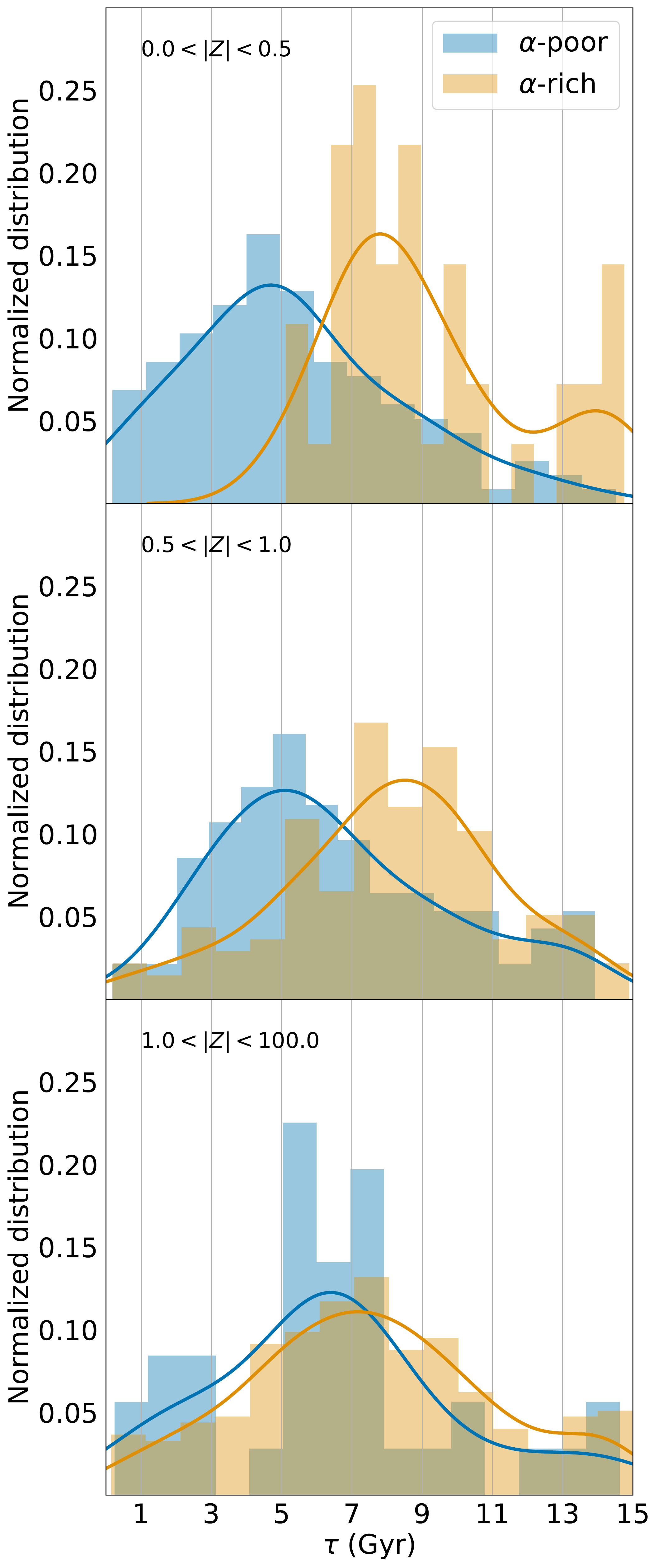}\end{center}
    \caption{Distributions of the age estimates for both the $\alpha$-poor and $\alpha$-rich of the \ktwo fields as a function of distance from the Galactic plane $|Z|$. While the median age of the $\alpha$-rich population is about constant as a function of vertical height, if not slightly decreasing with $|Z|$, the $\alpha$-poor population seems to increase in age as $|Z|$ increases, with the populations converging on an age of about 7 Gyr.  \label{fig:zdep}}
\end{figure}

It is also interesting to note that the ages of these two populations seem to converge as a function of height above the Galactic plane, a trend that is shown in Figure~\ref{fig:zdep}. \cite{haydenambre} finds similar trends, such that coeval populations of $\alpha$-rich and $\alpha$-poor stars are found to have the same vertical scale heights. 
These results potentially coincide with those from \cite{rendle+2019}, whose \ktwo Galactic archaeology results seem consistent with the "upside-down" formation model \citep[e.g.,][]{bird+2013}, wherein stars form in a vertically-extended disk, with subsequent formation occurring closer and closer to the Galactic plane.

We also tentatively recover the young, \al-rich stars seen in the \Kepler analysis of SA18 and the \ktwo analysis of \cite{rendle+2019}. The origin of these stars has not been definitively determined, but part of the population may be explained by stellar mergers.

To analyse the meaning of these results, it may be useful to place them within the context of the proposed scenarios in the literature for generating the bi-modal \al\ sequence, each of which comes with their own set of predictions concerning the relative ages of \al-rich and \al-poor populations and the homogeneity of these ages throughout the Galaxy.

One of these possible mechanisms is the radial migration of stars mixing together populations of different chemical origins. In this scenario, the radial migration of stars born at different radii in the disk results in the superposition of several chemical evolution sequences at any given location in the Galaxy. Stars with high [$\alpha$/Fe] at high [Fe/H], for example, could come from inner regions where more efficient star formation produced high metallicity gas relatively quickly. 
When mixed together, they form the sequence that we are familiar with. Therefore, the primary explanation for the lack of an age-metallicity relation is mostly attributed to stellar neighbors not having necessarily been born from the same gas, as stars have moved radially inwards and outwards in their orbits \citep[see e.g.][]{SB2009, Weinberg+2017, NBB2014, sharma+2020b}.
However, it is not entirely clear whether our results of similar \al-rich populations having strongly peaked ages in different parts of the Galaxy is consistent with this theory.

\begin{figure*}
	\plotone{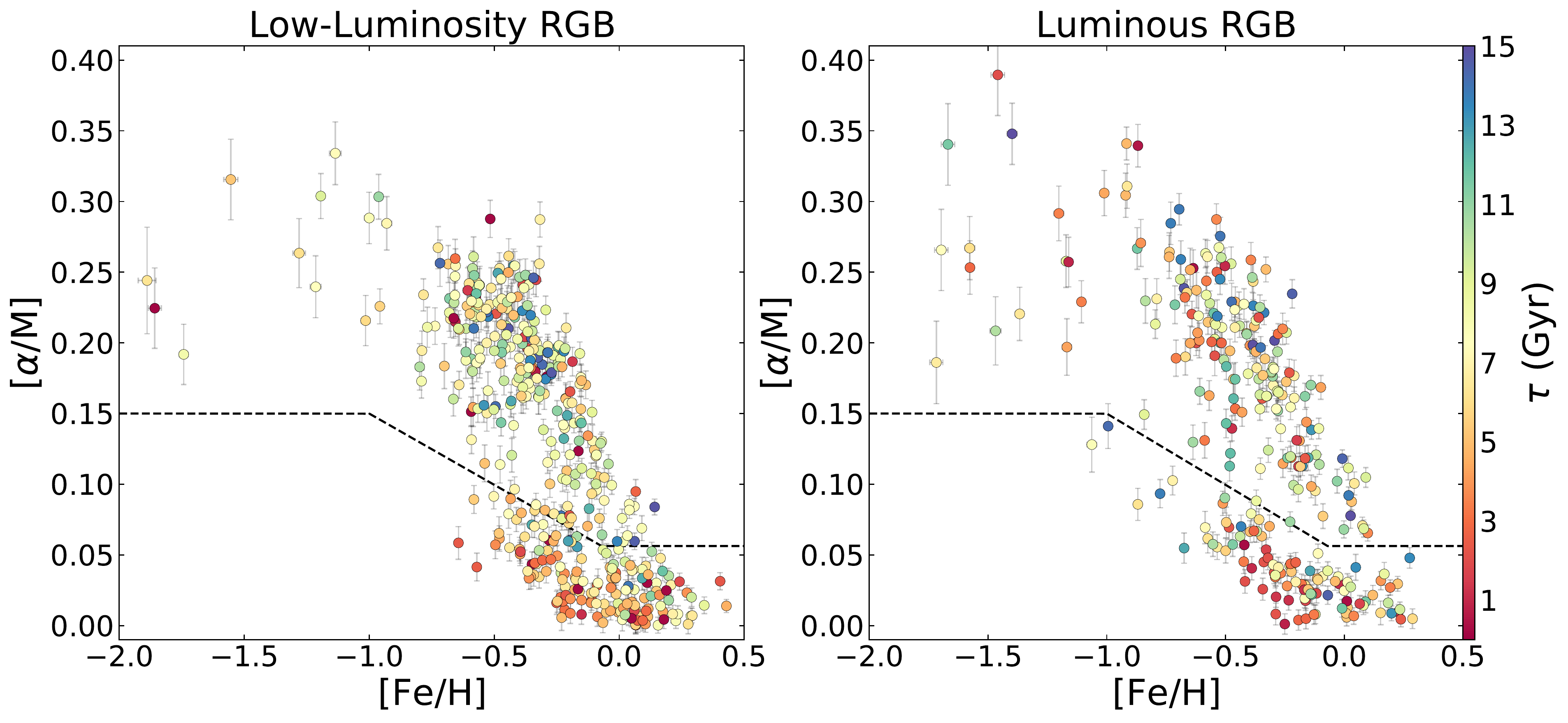}
	\caption{[\al/M] vs. [Fe/H] for the low-luminosity and luminous giants in our sample, colored by age. \label{fig:avm2}}
\end{figure*}
Another possible explanation is that the two loci of the bi-modal \al\ sequence are the products of two gas infall episodes that both spurred two independent periods of star formation. In this scenario, the \al-rich sequence was formed during a rapid infall episode that occurred about 10 billion years ago. This would be followed by a drought of star formation that would itself be followed by a gradual infall episode spurring star formation from about 8 billion years ago to present. The pristine gas in this second infall would dilute the disk gas, creating a new starting point at lower [Fe/H] for a second, less \al-enriched episode \citep{chiappini_matteucci_gratton1997}. Each of these star formation episodes would revive historic rates of SNe II, but have lesser effects on changing the rate of SNe Ia. \cite{Spitoni2019}, simulating this two-infall scenario, were able to successfully replicate the bi-modal \al\ sequence as well as the age trends found in SA18 for the high- and low-\al\  sequences. \cite{lian+2020} present a similar model which is able to better produce the density ridge-line between the \al-rich and \al-poor loci while also being able to more accurately reproduce the observed ages of \al-poor stars. Adding the \ktwo data, however, seems to suggest that this two-infall scenario would have to be at least slightly more complex if star formation happened at different times in different places in the Galaxy during the first, rapid episode. \cite{sharma+2020b} points out that the model by \cite{Spitoni2019} requires a loop in the tracks in the [\al/Fe] vs. [Fe/H] plane which does not agree with observations. Figure \ref{fig:avm2} shows the \ktwo sample in the [\al/Fe] vs. [Fe/H] plane with stars colored by their ages, but it is not clear that our results have the resolution nor the volume needed to fully explore the existence of this age loop.

A third proposed mechanism is that the bi-modal sequence is the product of star formation happening in clumpy bursts throughout the Galaxy. In this scenario, there is a background of star formation within the \al-poor sequence and the rest of the star formation takes place in gas-rich clumps that naturally arose in the disk. When star formation is spurred in these clumps, these populations of stars are initially enriched with \al\ elements as the number of SNe II is spurred by the formation of new, massive stars. Therefore, in this scenario, the \al-rich sequence is formed from a superposition of these clumpy star formation episodes \citep[see e.g.][]{Clarke2019}.
If assuming that the populations are not mixed together (by, for example, radial migration), this allows nearly identical chemical populations observed in different regions of the Galaxy to have diverging ages. This mechanism therefore seems promising as a potential explanation for why similar chemical populations in different parts of the Galaxy would have strongly peaked, discrepant ages. However, since the ages of the \al-rich populations in the three \ktwo fields seem to agree with each other but disagree the age of the \al-rich population in the \Kepler field, if this scenario is true, it is still curious why only the very local Galaxy seems to be unique in its star formation history.

Looking forward, the third data release of the \ktwo Galactic Archaeology Program will provide us with asteroseismic data for giants along fourteen more lines of sight in the Galaxy. Additionally, a similar analysis is possible with data from NASA's ongoing Transiting Exoplanet Survey Satellite mission. This data, along with the spectroscopy from the final APOGEE data releases and from other large-scale surveys, can be used to fill in a more complete picture of the age gradients for stellar populations throughout the Milky Way.

\acknowledgements

This paper is dedicated to the memory of Nikki Justice.

Funding for the Sloan Digital Sky Survey IV has been provided by the Alfred P. Sloan Foundation, the U.S. Department of Energy Office of Science, and the Participating Institutions. SDSS-IV acknowledges
support and resources from the Center for High-Performance Computing at
the University of Utah. The SDSS web site is www.sdss.org.

SDSS-IV is managed by the Astrophysical Research Consortium for the 
Participating Institutions of the SDSS Collaboration including the 
Brazilian Participation Group, the Carnegie Institution for Science, 
Carnegie Mellon University, the Chilean Participation Group, the French Participation Group, Harvard-Smithsonian Center for Astrophysics,
Instituto de Astrof\'isica de Canarias, The Johns Hopkins University, Kavli Institute for the Physics and Mathematics of the Universe (IPMU) / 
University of Tokyo, the Korean Participation Group, Lawrence Berkeley National Laboratory, 
Leibniz Institut f\"ur Astrophysik Potsdam (AIP),  
Max-Planck-Institut f\"ur Astronomie (MPIA Heidelberg), 
Max-Planck-Institut f\"ur Astrophysik (MPA Garching), 
Max-Planck-Institut f\"ur Extraterrestrische Physik (MPE), 
National Astronomical Observatories of China, New Mexico State University, 
New York University, University of Notre Dame, 
Observat\'ario Nacional / MCTI, The Ohio State University, 
Pennsylvania State University, Shanghai Astronomical Observatory, 
United Kingdom Participation Group,
Universidad Nacional Aut\'onoma de M\'exico, University of Arizona, 
University of Colorado Boulder, University of Oxford, University of Portsmouth, 
University of Utah, University of Virginia, University of Washington, University of Wisconsin, 
Vanderbilt University, and Yale University.

JTW and MHP acknowledge support from NASA Grant 80NSSC19K0115.
RAG acknowledges the support of the PLATO CNES grant.
SM would like to acknowledge support from the Spanish Ministry through the Ramon y Cajal fellowship number RYC-2015-17697.
DAGH acknowledges support from the State Research Agency (AEI) of the Spanish Ministry of Science, Innovation and Universities (MCIU) and the European Regional Development Fund (FEDER) under grant AYA2017-88254-P.


\bibliography{bibliography}{}

\begin{thebibliography}{}
\expandafter\ifx\csname natexlab\endcsname\relax\def\natexlab#1{#1}\fi
\providecommand{\url}[1]{\href{#1}{#1}}
\providecommand{\dodoi}[1]{doi:~\href{http://doi.org/#1}{\nolinkurl{#1}}}
\providecommand{\doeprint}[1]{\href{http://ascl.net/#1}{\nolinkurl{http://ascl.net/#1}}}
\providecommand{\doarXiv}[1]{\href{https://arxiv.org/abs/#1}{\nolinkurl{https://arxiv.org/abs/#1}}}

\bibitem[{{Ahumada} {et~al.}(2020){Ahumada}, {Prieto}, {Almeida}, {Anders},
  {Anderson}, {Andrews}, {Anguiano}, {Arcodia}, {Armengaud}, {Aubert}, {Avila},
  {Avila-Reese}, {Badenes}, {Balland }, {Barger}, {Barrera-Ballesteros},
  {Basu}, {Bautista}, {Beaton}, {Beers}, {Benavides}, {Bender}, {Bernardi},
  {Bershady}, {Beutler}, {Bidin}, {Bird}, {Bizyaev}, {Blanc}, {Blanton},
  {Boquien}, {Borissova}, {Bovy}, {Brand t}, {Brinkmann}, {Brownstein},
  {Bundy}, {Bureau}, {Burgasser}, {Burtin}, {Cano-D{\'\i}az}, {Capasso},
  {Cappellari}, {Carrera}, {Chabanier}, {Chaplin}, {Chapman}, {Cherinka},
  {Chiappini}, {Doohyun Choi}, {Chojnowski}, {Chung}, {Clerc}, {Coffey},
  {Comerford}, {Comparat}, {da Costa}, {Cousinou}, {Covey}, {Crane}, {Cunha},
  {Ilha}, {Dai}, {Damsted}, {Darling}, {Davidson}, {Davies}, {Dawson}, {De},
  {de la Macorra}, {De Lee}, {Queiroz}, {Deconto Machado}, {de la Torre},
  {Dell'Agli}, {du Mas des Bourboux}, {Diamond-Stanic}, {Dillon}, {Donor},
  {Drory}, {Duckworth}, {Dwelly}, {Ebelke}, {Eftekharzadeh}, {Davis Eigenbrot},
  {Elsworth}, {Eracleous}, {Erfanianfar}, {Escoffier}, {Fan}, {Farr},
  {Fern{\'a}ndez-Trincado}, {Feuillet}, {Finoguenov}, {Fofie},
  {Fraser-McKelvie}, {Frinchaboy}, {Fromenteau}, {Fu}, {Galbany}, {Garcia},
  {Garc{\'\i}a-Hern{\'a}ndez}, {Oehmichen}, {Ge}, {Maia}, {Geisler}, {Gelfand
  }, {Goddy}, {Gonzalez-Perez}, {Grabowski}, {Green}, {Grier}, {Guo}, {Guy},
  {Harding}, {Hasselquist}, {Hawken}, {Hayes}, {Hearty}, {Hekker}, {Hogg},
  {Holtzman}, {Horta}, {Hou}, {Hsieh}, {Huber}, {Hunt}, {Chitham}, {Imig},
  {Jaber}, {Angel}, {Johnson}, {Jones}, {J{\"o}nsson}, {Jullo}, {Kim},
  {Kinemuchi}, {Kirkpatrick}, {Kite}, {Klaene}, {Kneib}, {Kollmeier}, {Kong},
  {Kounkel}, {Krishnarao}, {Lacerna}, {Lan}, {Lane}, {Law}, {Le Goff}, {Leung},
  {Lewis}, {Li}, {Lian}, {Lin}, {Long}, {Longa-Pe{\~n}a}, {Lundgren}, {Lyke},
  {Ted Mackereth}, {MacLeod}, {Majewski}, {Manchado}, {Maraston}, {Martini},
  {Masseron}, {Masters}, {Mathur}, {McDermid}, {Merloni}, {Merrifield},
  {M{\'e}sz{\'a}ros}, {Miglio}, {Minniti}, {Minsley}, {Miyaji}, {Mohammad},
  {Mosser}, {Mueller}, {Muna}, {Mu{\~n}oz-Guti{\'e}rrez}, {Myers}, {Nadathur},
  {Nair}, {Nandra}, {do Nascimento}, {Nevin}, {Newman}, {Nidever}, {Nitschelm},
  {Noterdaeme}, {O'Connell}, {Olmstead}, {Oravetz}, {Oravetz}, {Osorio},
  {Pace}, {Padilla}, {Palanque-Delabrouille}, {Palicio}, {Pan}, {Pan},
  {Parker}, {Paviot}, {Peirani}, {Ram{\'r}ez}, {Penny}, {Percival},
  {Perez-Fournon}, {P{\'e}rez-R{\`a}fols}, {Petitjean}, {Pieri},
  {Pinsonneault}, {Poovelil}, {Povick}, {Prakash}, {Price-Whelan}, {Raddick},
  {Raichoor}, {Ray}, {Rembold}, {Rezaie}, {Riffel}, {Riffel}, {Rix}, {Robin},
  {Roman-Lopes}, {Rom{\'a}n-Z{\'u}{\~n}iga}, {Rose}, {Ross}, {Rossi}, {Rowland
  s}, {Rubin}, {Salvato}, {S{\'a}nchez}, {S{\'a}nchez-Menguiano},
  {S{\'a}nchez-Gallego}, {Sayres}, {Schaefer}, {Schiavon}, {Schimoia},
  {Schlafly}, {Schlegel}, {Schneider}, {Schultheis}, {Schwope}, {Seo},
  {Serenelli}, {Shafieloo}, {Shamsi}, {Shao}, {Shen}, {Shetrone}, {Shirley},
  {Aguirre}, {Simon}, {Skrutskie}, {Slosar}, {Smethurst}, {Sobeck}, {Sodi},
  {Souto}, {Stark}, {Stassun}, {Steinmetz}, {Stello}, {Stermer},
  {Storchi-Bergmann}, {Streblyanska}, {Stringfellow}, {Stutz}, {Su{\'a}rez},
  {Sun}, {Taghizadeh-Popp}, {Talbot}, {Tayar}, {Thakar}, {Theriault}, {Thomas},
  {Thomas}, {Tinker}, {Tojeiro}, {Toledo}, {Tremonti}, {Troup}, {Tuttle},
  {Unda-Sanzana}, {Valentini}, {Vargas-Gonz{\'a}lez}, {Vargas-Maga{\~n}a},
  {V{\'a}zquez-Mata}, {Vivek}, {Wake}, {Wang}, {Weaver}, {Weijmans}, {Wild},
  {Wilson}, {Wilson}, {Wolthuis}, {Wood-Vasey}, {Yan}, {Yang}, {Y{\`e}che},
  {Zamora}, {Zarrouk}, {Zasowski}, {Zhang}, {Zhao}, {Zhao}, {Zheng}, {Zheng},
  {Zhu}, \& {Zou}}]{apogeedr16}
{Ahumada}, R., {Prieto}, C.~A., {Almeida}, A., {et~al.} 2020, \apjs, 249, 3

\bibitem[{{Aller} \& {Greenstein}(1960)}]{aller1960}
{Aller}, L.~H., \& {Greenstein}, J.~L. 1960, \apjs, 5, 139

\bibitem[{{Anders} {et~al.}(2017){Anders}, {Chiappini}, {Rodrigues}, {Miglio},
  {Montalb{\'a}n}, {Mosser}, {Girardi}, {Valentini}, {Noels}, {Morel},
  {Johnson}, {Schultheis}, {Baudin}, {de Assis Peralta}, {Hekker},
  {Theme{\ss}l}, {Kallinger}, {Garc{\'\i}a}, {Mathur}, {Baglin}, {Santiago},
  {Martig}, {Minchev}, {Steinmetz}, {da Costa}, {Maia}, {Allende Prieto},
  {Cunha}, {Beers}, {Epstein}, {Garc{\'\i}a P{\'e}rez},
  {Garc{\'\i}a-Hern{\'a}ndez}, {Harding}, {Holtzman}, {Majewski},
  {M{\'e}sz{\'a}ros}, {Nidever}, {Pan}, {Pinsonneault}, {Schiavon},
  {Schneider}, {Shetrone}, {Stassun}, {Zamora}, \& {Zasowski}}]{corogee}
{Anders}, F., {Chiappini}, C., {Rodrigues}, T.~S., {et~al.} 2017, \aap, 597,
  A30

\bibitem[{{Astropy Collaboration} {et~al.}(2013){Astropy Collaboration},
  {Robitaille}, {Tollerud}, {Greenfield}, {Droettboom}, {Bray}, {Aldcroft},
  {Davis}, {Ginsburg}, {Price-Whelan}, {Kerzendorf}, {Conley}, {Crighton},
  {Barbary}, {Muna}, {Ferguson}, {Grollier}, {Parikh}, {Nair}, {Unther},
  {Deil}, {Woillez}, {Conseil}, {Kramer}, {Turner}, {Singer}, {Fox}, {Weaver},
  {Zabalza}, {Edwards}, {Azalee Bostroem}, {Burke}, {Casey}, {Crawford},
  {Dencheva}, {Ely}, {Jenness}, {Labrie}, {Lim}, {Pierfederici}, {Pontzen},
  {Ptak}, {Refsdal}, {Servillat}, \& {Streicher}}]{astropy:2013}
{Astropy Collaboration}, {Robitaille}, T.~P., {Tollerud}, E.~J., {et~al.} 2013,
  \aap, 558, A33

\bibitem[{{Bailer-Jones}(2015)}]{BailerJones2015}
{Bailer-Jones}, C. A.~L. 2015, \pasp, 127, 994

\bibitem[{{Bedding} {et~al.}(2011){Bedding}, {Mosser}, {Huber},
  {Montalb{\'a}n}, {Beck}, {Christensen-Dalsgaard}, {Elsworth}, {Garc{\'\i}a},
  {Miglio}, {Stello}, {White}, {De Ridder}, {Hekker}, {Aerts}, {Barban},
  {Belkacem}, {Broomhall}, {Brown}, {Buzasi}, {Carrier}, {Chaplin}, {di Mauro},
  {Dupret}, {Frandsen}, {Gilliland }, {Goupil}, {Jenkins}, {Kallinger},
  {Kawaler}, {Kjeldsen}, {Mathur}, {Noels}, {Silva Aguirre}, \&
  {Ventura}}]{bedding+2011}
{Bedding}, T.~R., {Mosser}, B., {Huber}, D., {et~al.} 2011, \nat, 471, 608

\bibitem[{{Bensby} {et~al.}(2003){Bensby}, {Feltzing}, \&
  {Lundstr{\"o}m}}]{Bensby2003}
{Bensby}, T., {Feltzing}, S., \& {Lundstr{\"o}m}, I. 2003, \aap, 410, 527

\bibitem[{{Bird} {et~al.}(2013){Bird}, {Kazantzidis}, {Weinberg}, {Guedes},
  {Callegari}, {Mayer}, \& {Madau}}]{bird+2013}
{Bird}, J.~C., {Kazantzidis}, S., {Weinberg}, D.~H., {et~al.} 2013, \apj, 773,
  43

\bibitem[{{Blanton} {et~al.}(2017){Blanton}, {Bershady}, {Abolfathi},
  {Albareti}, {Allende Prieto}, {Almeida}, {Alonso-Garc{\'\i}a}, {Anders},
  {Anderson}, {Andrews}, {Aquino-Ort{\'\i}z}, {Arag{\'o}n-Salamanca},
  {Argudo-Fern{\'a}ndez}, {Armengaud}, {Aubourg}, {Avila-Reese}, {Badenes},
  {Bailey}, {Barger}, {Barrera-Ballesteros}, {Bartosz}, {Bates}, {Baumgarten},
  {Bautista}, {Beaton}, {Beers}, {Belfiore}, {Bender}, {Berlind}, {Bernardi},
  {Beutler}, {Bird}, {Bizyaev}, {Blanc}, {Blomqvist}, {Bolton}, {Boquien},
  {Borissova}, {van den Bosch}, {Bovy}, {Brandt}, {Brinkmann}, {Brownstein},
  {Bundy}, {Burgasser}, {Burtin}, {Busca}, {Cappellari}, {Delgado Carigi},
  {Carlberg}, {Carnero Rosell}, {Carrera}, {Chanover}, {Cherinka}, {Cheung},
  {G{\'o}mez Maqueo Chew}, {Chiappini}, {Choi}, {Chojnowski}, {Chuang},
  {Chung}, {Cirolini}, {Clerc}, {Cohen}, {Comparat}, {da Costa}, {Cousinou},
  {Covey}, {Crane}, {Croft}, {Cruz-Gonzalez}, {Garrido Cuadra}, {Cunha},
  {Damke}, {Darling}, {Davies}, {Dawson}, {de la Macorra}, {Dell'Agli}, {De
  Lee}, {Delubac}, {Di Mille}, {Diamond-Stanic}, {Cano-D{\'\i}az}, {Donor},
  {Downes}, {Drory}, {du Mas des Bourboux}, {Duckworth}, {Dwelly}, {Dyer},
  {Ebelke}, {Eigenbrot}, {Eisenstein}, {Emsellem}, {Eracleous}, {Escoffier},
  {Evans}, {Fan}, {Fern{\'a}ndez-Alvar}, {Fernandez-Trincado}, {Feuillet},
  {Finoguenov}, {Fleming}, {Font-Ribera}, {Fredrickson}, {Freischlad},
  {Frinchaboy}, {Fuentes}, {Galbany}, {Garcia-Dias},
  {Garc{\'\i}a-Hern{\'a}ndez}, {Gaulme}, {Geisler}, {Gelfand},
  {Gil-Mar{\'\i}n}, {Gillespie}, {Goddard}, {Gonzalez-Perez}, {Grabowski},
  {Green}, {Grier}, {Gunn}, {Guo}, {Guy}, {Hagen}, {Hahn}, {Hall}, {Harding},
  {Hasselquist}, {Hawley}, {Hearty}, {Gonzalez Hern{\'a}ndez}, {Ho}, {Hogg},
  {Holley-Bockelmann}, {Holtzman}, {Holzer}, {Huehnerhoff}, {Hutchinson},
  {Hwang}, {Ibarra-Medel}, {da Silva Ilha}, {Ivans}, {Ivory}, {Jackson},
  {Jensen}, {Johnson}, {Jones}, {J{\"o}nsson}, {Jullo}, {Kamble}, {Kinemuchi},
  {Kirkby}, {Kitaura}, {Klaene}, {Knapp}, {Kneib}, {Kollmeier}, {Lacerna},
  {Lane}, {Lang}, {Law}, {Lazarz}, {Lee}, {Le Goff}, {Liang}, {Li}, {Li},
  {Lian}, {Lima}, {Lin}, {Lin}, {Bertran de Lis}, {Liu}, {de Icaza Lizaola},
  {Long}, {Lucatello}, {Lundgren}, {MacDonald}, {Deconto Machado}, {MacLeod},
  {Mahadevan}, {Geimba Maia}, {Maiolino}, {Majewski}, {Malanushenko},
  {Malanushenko}, {Manchado}, {Mao}, {Maraston}, {Marques-Chaves}, {Masseron},
  {Masters}, {McBride}, {McDermid}, {McGrath}, {McGreer}, {Medina Pe{\~n}a},
  {Melendez}, {Merloni}, {Merrifield}, {Meszaros}, {Meza}, {Minchev},
  {Minniti}, {Miyaji}, {More}, {Mulchaey}, {M{\"u}ller-S{\'a}nchez}, {Muna},
  {Munoz}, {Myers}, {Nair}, {Nandra}, {Correa do Nascimento}, {Negrete},
  {Ness}, {Newman}, {Nichol}, {Nidever}, {Nitschelm}, {Ntelis}, {O'Connell},
  {Oelkers}, {Oravetz}, {Oravetz}, {Pace}, {Padilla}, {Palanque-Delabrouille},
  {Alonso Palicio}, {Pan}, {Parejko}, {Parikh}, {P{\^a}ris}, {Park}, {Patten},
  {Peirani}, {Pellejero-Ibanez}, {Penny}, {Percival}, {Perez-Fournon},
  {Petitjean}, {Pieri}, {Pinsonneault}, {Pisani}, {Poleski}, {Prada},
  {Prakash}, {Queiroz}, {Raddick}, {Raichoor}, {Barboza Rembold}, {Richstein},
  {Riffel}, {Riffel}, {Rix}, {Robin}, {Rockosi}, {Rodr{\'\i}guez-Torres},
  {Roman-Lopes}, {Rom{\'a}n-Z{\'u}{\~n}iga}, {Rosado}, {Ross}, {Rossi}, {Ruan},
  {Ruggeri}, {Rykoff}, {Salazar-Albornoz}, {Salvato}, {S{\'a}nchez}, {Aguado},
  {S{\'a}nchez-Gallego}, {Santana}, {Santiago}, {Sayres}, {Schiavon}, {da Silva
  Schimoia}, {Schlafly}, {Schlegel}, {Schneider}, {Schultheis}, {Schuster},
  {Schwope}, {Seo}, {Shao}, {Shen}, {Shetrone}, {Shull}, {Simon}, {Skinner},
  {Skrutskie}, {Slosar}, {Smith}, {Sobeck}, {Sobreira}, {Somers}, {Souto},
  {Stark}, {Stassun}, {Stauffer}, {Steinmetz}, {Storchi-Bergmann},
  {Streblyanska}, {Stringfellow}, {Su{\'a}rez}, {Sun}, {Suzuki}, {Szigeti},
  {Taghizadeh-Popp}, {Tang}, {Tao}, {Tayar}, {Tembe}, {Teske}, {Thakar},
  {Thomas}, {Thompson}, {Tinker}, {Tissera}, {Tojeiro}, {Hernandez Toledo}, {de
  la Torre}, {Tremonti}, {Troup}, {Valenzuela}, {Martinez Valpuesta},
  {Vargas-Gonz{\'a}lez}, {Vargas-Maga{\~n}a}, {Vazquez}, {Villanova}, {Vivek},
  {Vogt}, {Wake}, {Walterbos}, {Wang}, {Weaver}, {Weijmans}, {Weinberg},
  {Westfall}, {Whelan}, {Wild}, {Wilson}, {Wood-Vasey}, {Wylezalek}, {Xiao},
  {Yan}, {Yang}, {Ybarra}, {Y{\`e}che}, {Zakamska}, {Zamora}, {Zarrouk},
  {Zasowski}, {Zhang}, {Zhao}, {Zheng}, {Zheng}, {Zhou}, {Zhou}, {Zhu},
  {Zoccali}, \& {Zou}}]{BlantonAndBershady2017}
{Blanton}, M.~R., {Bershady}, M.~A., {Abolfathi}, B., {et~al.} 2017, \aj, 154,
  28

\bibitem[{{Bovy} {et~al.}(2014){Bovy}, {Nidever}, {Rix}, {Girardi}, {Zasowski},
  {Chojnowski}, {Holtzman}, {Epstein}, {Frinchaboy}, {Hayden}, {Rodrigues},
  {Majewski}, {Johnson}, {Pinsonneault}, {Stello}, {Allende Prieto}, {Andrews},
  {Basu}, {Beers}, {Bizyaev}, {Burton}, {Chaplin}, {Cunha}, {Elsworth},
  {Garc{\'\i}a}, {Garc{\'\i}a-Her{\'n}andez}, {Garc{\'\i}a P{\'e}rez},
  {Hearty}, {Hekker}, {Kallinger}, {Kinemuchi}, {Koesterke},
  {M{\'e}sz{\'a}ros}, {Mosser}, {O'Connell}, {Oravetz}, {Pan}, {Robin},
  {Schiavon}, {Schneider}, {Schultheis}, {Serenelli}, {Shetrone}, {Silva
  Aguirre}, {Simmons}, {Skrutskie}, {Smith}, {Stassun}, {Weinberg}, {Wilson},
  \& {Zamora}}]{Bovy2014}
{Bovy}, J., {Nidever}, D.~L., {Rix}, H.-W., {et~al.} 2014, \apj, 790, 127

\bibitem[{Bowen \& Vaughan(1973)}]{bowen73}
Bowen, I.~S., \& Vaughan, A.~H. 1973, Appl. Opt., 12, 1430

\bibitem[{{Brown} {et~al.}(1991){Brown}, {Gilliland}, {Noyes}, \&
  {Ramsey}}]{brown1991}
{Brown}, T.~M., {Gilliland}, R.~L., {Noyes}, R.~W., \& {Ramsey}, L.~W. 1991,
  \apj, 368, 599

\bibitem[{{Casagrande} {et~al.}(2016){Casagrande}, {Silva Aguirre},
  {Schlesinger}, {Stello}, {Huber}, {Serenelli}, {Sch{\"o}nrich}, {Cassisi},
  {Pietrinferni}, {Hodgkin}, {Milone}, {Feltzing}, \&
  {Asplund}}]{casagrande+2016}
{Casagrande}, L., {Silva Aguirre}, V., {Schlesinger}, K.~J., {et~al.} 2016,
  \mnras, 455, 987

\bibitem[{{Chaplin} {et~al.}(2008){Chaplin}, {Houdek}, {Appourchaux},
  {Elsworth}, {New}, \& {Toutain}}]{chaplin+2008}
{Chaplin}, W.~J., {Houdek}, G., {Appourchaux}, T., {et~al.} 2008, \aap, 485,
  813

\bibitem[{{Chiappini} {et~al.}(1997){Chiappini}, {Matteucci}, \&
  {Gratton}}]{chiappini_matteucci_gratton1997}
{Chiappini}, C., {Matteucci}, F., \& {Gratton}, R. 1997, \apj, 477, 765

\bibitem[{{Chiappini} {et~al.}(2015){Chiappini}, {Anders}, {Rodrigues},
  {Miglio}, {Montalb{\'a}n}, {Mosser}, {Girardi}, {Valentini}, {Noels},
  {Morel}, {Minchev}, {Steinmetz}, {Santiago}, {Schultheis}, {Martig}, {da
  Costa}, {Maia}, {Allende Prieto}, {de Assis Peralta}, {Hekker},
  {Theme{\ss}l}, {Kallinger}, {Garc{\'\i}a}, {Mathur}, {Baudin}, {Beers},
  {Cunha}, {Harding}, {Holtzman}, {Majewski}, {M{\'e}sz{\'a}ros}, {Nidever},
  {Pan}, {Schiavon}, {Shetrone}, {Schneider}, \& {Stassun}}]{Chiappini2015}
{Chiappini}, C., {Anders}, F., {Rodrigues}, T.~S., {et~al.} 2015, \aap, 576,
  L12

\bibitem[{{Clarke} {et~al.}(2019){Clarke}, {Debattista}, {Nidever}, {Loebman},
  {Simons}, {Kassin}, {Du}, {Ness}, {Fisher}, {Quinn}, {Wadsley}, {Freeman}, \&
  {Popescu}}]{Clarke2019}
{Clarke}, A.~J., {Debattista}, V.~P., {Nidever}, D.~L., {et~al.} 2019, \mnras,
  484, 3476

\bibitem[{{da Silva} {et~al.}(2006){da Silva}, {Girardi}, {Pasquini},
  {Setiawan}, {von der L{\"u}he}, {de Medeiros}, {Hatzes}, {D{\"o}llinger}, \&
  {Weiss}}]{param1}
{da Silva}, L., {Girardi}, L., {Pasquini}, L., {et~al.} 2006, \aap, 458, 609

\bibitem[{{Elsworth} {et~al.}(2020){Elsworth}, {Theme{\ss}l}, {Hekker}, \&
  {Chaplin}}]{ElsworthBHM}
{Elsworth}, Y., {Theme{\ss}l}, N., {Hekker}, S., \& {Chaplin}, W. 2020,
  Research Notes of the American Astronomical Society, 4, 177

\bibitem[{{Elsworth} {et~al.}(2019){Elsworth}, {Hekker}, {Johnson},
  {Kallinger}, {Mosser}, {Pinsonneault}, {Hon}, {Kuszlewicz}, {Miglio},
  {Serenelli}, {Stello}, {Tayar}, \& {Vrard}}]{apokascstates}
{Elsworth}, Y., {Hekker}, S., {Johnson}, J.~A., {et~al.} 2019, \mnras, 489,
  4641

\bibitem[{{Epstein} {et~al.}(2014){Epstein}, {Elsworth}, {Johnson}, {Shetrone},
  {Mosser}, {Hekker}, {Tayar}, {Harding}, {Pinsonneault}, {Silva Aguirre},
  {Basu}, {Beers}, {Bizyaev}, {Bedding}, {Chaplin}, {Frinchaboy},
  {Garc{\'\i}a}, {Garc{\'\i}a P{\'e}rez}, {Hearty}, {Huber}, {Ivans},
  {Majewski}, {Mathur}, {Nidever}, {Serenelli}, {Schiavon}, {Schneider},
  {Sch{\"o}nrich}, {Sobeck}, {Stassun}, {Stello}, \& {Zasowski}}]{Epstein2014}
{Epstein}, C.~R., {Elsworth}, Y.~P., {Johnson}, J.~A., {et~al.} 2014, \apjl,
  785, L28

\bibitem[{{Fuhrmann}(1998)}]{fuhrmann1998}
{Fuhrmann}, K. 1998, \aap, 338, 161

\bibitem[{{Fuhrmann}(2011)}]{fuhrmann2011}
---. 2011, \mnras, 414, 2893

\bibitem[{{Gaia Collaboration} {et~al.}(2016){Gaia Collaboration}, {Prusti},
  {de Bruijne}, {Brown}, {Vallenari}, {Babusiaux}, {Bailer-Jones}, {Bastian},
  {Biermann}, {Evans}, {Eyer}, {Jansen}, {Jordi}, {Klioner}, {Lammers},
  {Lindegren}, {Luri}, {Mignard}, {Milligan}, {Panem}, {Poinsignon},
  {Pourbaix}, {Randich}, {Sarri}, {Sartoretti}, {Siddiqui}, {Soubiran},
  {Valette}, {van Leeuwen}, {Walton}, {Aerts}, {Arenou}, {Cropper}, {Drimmel},
  {H{\o}g}, {Katz}, {Lattanzi}, {O'Mullane}, {Grebel}, {Holland}, {Huc},
  {Passot}, {Bramante}, {Cacciari}, {Casta{\~n}eda}, {Chaoul}, {Cheek}, {De
  Angeli}, {Fabricius}, {Guerra}, {Hern{\'a}ndez}, {Jean-Antoine-Piccolo},
  {Masana}, {Messineo}, {Mowlavi}, {Nienartowicz}, {Ord{\'o}{\~n}ez-Blanco},
  {Panuzzo}, {Portell}, {Richards}, {Riello}, {Seabroke}, {Tanga},
  {Th{\'e}venin}, {Torra}, {Els}, {Gracia-Abril}, {Comoretto},
  {Garcia-Reinaldos}, {Lock}, {Mercier}, {Altmann}, {Andrae}, {Astraatmadja},
  {Bellas-Velidis}, {Benson}, {Berthier}, {Blomme}, {Busso}, {Carry},
  {Cellino}, {Clementini}, {Cowell}, {Creevey}, {Cuypers}, {Davidson}, {De
  Ridder}, {de Torres}, {Delchambre}, {Dell'Oro}, {Ducourant}, {Fr{\'e}mat},
  {Garc{\'\i}a-Torres}, {Gosset}, {Halbwachs}, {Hambly}, {Harrison}, {Hauser},
  {Hestroffer}, {Hodgkin}, {Huckle}, {Hutton}, {Jasniewicz}, {Jordan},
  {Kontizas}, {Korn}, {Lanzafame}, {Manteiga}, {Moitinho}, {Muinonen},
  {Osinde}, {Pancino}, {Pauwels}, {Petit}, {Recio-Blanco}, {Robin}, {Sarro},
  {Siopis}, {Smith}, {Smith}, {Sozzetti}, {Thuillot}, {van Reeven}, {Viala},
  {Abbas}, {Abreu Aramburu}, {Accart}, {Aguado}, {Allan}, {Allasia},
  {Altavilla}, {{\'A}lvarez}, {Alves}, {Anderson}, {Andrei}, {Anglada Varela},
  {Antiche}, {Antoja}, {Ant{\'o}n}, {Arcay}, {Atzei}, {Ayache}, {Bach},
  {Baker}, {Balaguer-N{\'u}{\~n}ez}, {Barache}, {Barata}, {Barbier}, {Barblan},
  {Baroni}, {Barrado y Navascu{\'e}s}, {Barros}, {Barstow}, {Becciani},
  {Bellazzini}, {Bellei}, {Bello Garc{\'\i}a}, {Belokurov}, {Bendjoya},
  {Berihuete}, {Bianchi}, {Bienaym{\'e}}, {Billebaud}, {Blagorodnova},
  {Blanco-Cuaresma}, {Boch}, {Bombrun}, {Borrachero}, {Bouquillon}, {Bourda},
  {Bouy}, {Bragaglia}, {Breddels}, {Brouillet}, {Br{\"u}semeister},
  {Bucciarelli}, {Budnik}, {Burgess}, {Burgon}, {Burlacu}, {Busonero}, {Buzzi},
  {Caffau}, {Cambras}, {Campbell}, {Cancelliere}, {Cantat-Gaudin}, {Carlucci},
  {Carrasco}, {Castellani}, {Charlot}, {Charnas}, {Charvet}, {Chassat},
  {Chiavassa}, {Clotet}, {Cocozza}, {Collins}, {Collins}, {Costigan}, {Crifo},
  {Cross}, {Crosta}, {Crowley}, {Dafonte}, {Damerdji}, {Dapergolas}, {David},
  {David}, {De Cat}, {de Felice}, {de Laverny}, {De Luise}, {De March}, {de
  Martino}, {de Souza}, {Debosscher}, {del Pozo}, {Delbo}, {Delgado},
  {Delgado}, {di Marco}, {Di Matteo}, {Diakite}, {Distefano}, {Dolding}, {Dos
  Anjos}, {Drazinos}, {Dur{\'a}n}, {Dzigan}, {Ecale}, {Edvardsson}, {Enke},
  {Erdmann}, {Escolar}, {Espina}, {Evans}, {Eynard Bontemps}, {Fabre},
  {Fabrizio}, {Faigler}, {Falc{\~a}o}, {Farr{\`a}s Casas}, {Faye}, {Federici},
  {Fedorets}, {Fern{\'a}ndez-Hern{\'a}ndez}, {Fernique}, {Fienga}, {Figueras},
  {Filippi}, {Findeisen}, {Fonti}, {Fouesneau}, {Fraile}, {Fraser}, {Fuchs},
  {Furnell}, {Gai}, {Galleti}, {Galluccio}, {Garabato}, {Garc{\'\i}a-Sedano},
  {Gar{\'e}}, {Garofalo}, {Garralda}, {Gavras}, {Gerssen}, {Geyer}, {Gilmore},
  {Girona}, {Giuffrida}, {Gomes}, {Gonz{\'a}lez-Marcos},
  {Gonz{\'a}lez-N{\'u}{\~n}ez}, {Gonz{\'a}lez-Vidal}, {Granvik}, {Guerrier},
  {Guillout}, {Guiraud}, {G{\'u}rpide}, {Guti{\'e}rrez-S{\'a}nchez}, {Guy},
  {Haigron}, {Hatzidimitriou}, {Haywood}, {Heiter}, {Helmi}, {Hobbs},
  {Hofmann}, {Holl}, {Holland }, {Hunt}, {Hypki}, {Icardi}, {Irwin}, {Jevardat
  de Fombelle}, {Jofr{\'e}}, {Jonker}, {Jorissen}, {Julbe}, {Karampelas},
  {Kochoska}, {Kohley}, {Kolenberg}, {Kontizas}, {Koposov}, {Kordopatis},
  {Koubsky}, {Kowalczyk}, {Krone-Martins}, {Kudryashova}, {Kull}, {Bachchan},
  {Lacoste-Seris}, {Lanza}, {Lavigne}, {Le Poncin-Lafitte}, {Lebreton},
  {Lebzelter}, {Leccia}, {Leclerc}, {Lecoeur-Taibi}, {Lemaitre}, {Lenhardt},
  {Leroux}, {Liao}, {Licata}, {Lindstr{\o}m}, {Lister}, {Livanou}, {Lobel},
  {L{\"o}ffler}, {L{\'o}pez}, {Lopez-Lozano}, {Lorenz}, {Loureiro},
  {MacDonald}, {Magalh{\~a}es Fernandes}, {Managau}, {Mann}, {Mantelet},
  {Marchal}, {Marchant}, {Marconi}, {Marie}, {Marinoni}, {Marrese},
  {Marschalk{\'o}}, {Marshall}, {Mart{\'\i}n-Fleitas}, {Martino}, {Mary},
  {Matijevi{\v{c}}}, {Mazeh}, {McMillan}, {Messina}, {Mestre}, {Michalik},
  {Millar}, {Miranda}, {Molina}, {Molinaro}, {Molinaro}, {Moln{\'a}r},
  {Moniez}, {Montegriffo}, {Monteiro}, {Mor}, {Mora}, {Morbidelli}, {Morel},
  {Morgenthaler}, {Morley}, {Morris}, {Mulone}, {Muraveva}, {Musella},
  {Narbonne}, {Nelemans}, {Nicastro}, {Noval}, {Ord{\'e}novic},
  {Ordieres-Mer{\'e}}, {Osborne}, {Pagani}, {Pagano}, {Pailler}, {Palacin},
  {Palaversa}, {Parsons}, {Paulsen}, {Pecoraro}, {Pedrosa}, {Pentik{\"a}inen},
  {Pereira}, {Pichon}, {Piersimoni}, {Pineau}, {Plachy}, {Plum}, {Poujoulet},
  {Pr{\v{s}}a}, {Pulone}, {Ragaini}, {Rago}, {Rambaux}, {Ramos-Lerate},
  {Ranalli}, {Rauw}, {Read}, {Regibo}, {Renk}, {Reyl{\'e}}, {Ribeiro},
  {Rimoldini}, {Ripepi}, {Riva}, {Rixon}, {Roelens}, {Romero-G{\'o}mez},
  {Rowell}, {Royer}, {Rudolph}, {Ruiz-Dern}, {Sadowski}, {Sagrist{\`a}
  Sell{\'e}s}, {Sahlmann}, {Salgado}, {Salguero}, {Sarasso}, {Savietto},
  {Schnorhk}, {Schultheis}, {Sciacca}, {Segol}, {Segovia}, {Segransan},
  {Serpell}, {Shih}, {Smareglia}, {Smart}, {Smith}, {Solano}, {Solitro},
  {Sordo}, {Soria Nieto}, {Souchay}, {Spagna}, {Spoto}, {Stampa}, {Steele},
  {Steidelm{\"u}ller}, {Stephenson}, {Stoev}, {Suess}, {S{\"u}veges}, {Surdej},
  {Szabados}, {Szegedi-Elek}, {Tapiador}, {Taris}, {Tauran}, {Taylor},
  {Teixeira}, {Terrett}, {Tingley}, {Trager}, {Turon}, {Ulla}, {Utrilla},
  {Valentini}, {van Elteren}, {Van Hemelryck}, {van Leeuwen}, {Varadi},
  {Vecchiato}, {Veljanoski}, {Via}, {Vicente}, {Vogt}, {Voss}, {Votruba},
  {Voutsinas}, {Walmsley}, {Weiler}, {Weingrill}, {Werner}, {Wevers},
  {Whitehead}, {Wyrzykowski}, {Yoldas}, {{\v{Z}}erjal}, {Zucker}, {Zurbach},
  {Zwitter}, {Alecu}, {Allen}, {Allende Prieto}, {Amorim},
  {Anglada-Escud{\'e}}, {Arsenijevic}, {Azaz}, {Balm}, {Beck}, {Bernstein},
  {Bigot}, {Bijaoui}, {Blasco}, {Bonfigli}, {Bono}, {Boudreault}, {Bressan},
  {Brown}, {Brunet}, {Bunclark}, {Buonanno}, {Butkevich}, {Carret}, {Carrion},
  {Chemin}, {Ch{\'e}reau}, {Corcione}, {Darmigny}, {de Boer}, {de Teodoro}, {de
  Zeeuw}, {Delle Luche}, {Domingues}, {Dubath}, {Fodor}, {Fr{\'e}zouls},
  {Fries}, {Fustes}, {Fyfe}, {Gallardo}, {Gallegos}, {Gardiol}, {Gebran},
  {Gomboc}, {G{\'o}mez}, {Grux}, {Gueguen}, {Heyrovsky}, {Hoar}, {Iannicola},
  {Isasi Parache}, {Janotto}, {Joliet}, {Jonckheere}, {Keil}, {Kim},
  {Klagyivik}, {Klar}, {Knude}, {Kochukhov}, {Kolka}, {Kos}, {Kutka}, {Lainey},
  {LeBouquin}, {Liu}, {Loreggia}, {Makarov}, {Marseille}, {Martayan},
  {Martinez-Rubi}, {Massart}, {Meynadier}, {Mignot}, {Munari}, {Nguyen},
  {Nordlander}, {Ocvirk}, {O'Flaherty}, {Olias Sanz}, {Ortiz}, {Osorio},
  {Oszkiewicz}, {Ouzounis}, {Palmer}, {Park}, {Pasquato}, {Peltzer}, {Peralta},
  {P{\'e}turaud}, {Pieniluoma}, {Pigozzi}, {Poels}, {Prat}, {Prod'homme},
  {Raison}, {Rebordao}, {Risquez}, {Rocca-Volmerange}, {Rosen}, {Ruiz-Fuertes},
  {Russo}, {Sembay}, {Serraller Vizcaino}, {Short}, {Siebert}, {Silva},
  {Sinachopoulos}, {Slezak}, {Soffel}, {Sosnowska}, {Strai{\v{z}}ys}, {ter
  Linden}, {Terrell}, {Theil}, {Tiede}, {Troisi}, {Tsalmantza}, {Tur},
  {Vaccari}, {Vachier}, {Valles}, {Van Hamme}, {Veltz}, {Virtanen}, {Wallut},
  {Wichmann}, {Wilkinson}, {Ziaeepour}, \& {Zschocke}}]{gaia1}
{Gaia Collaboration}, {Prusti}, T., {de Bruijne}, J.~H.~J., {et~al.} 2016,
  \aap, 595, A1

\bibitem[{{Gaia Collaboration} {et~al.}(2018){Gaia Collaboration}, {Brown},
  {Vallenari}, {Prusti}, {de Bruijne}, {Babusiaux}, {Bailer-Jones}, {Biermann},
  {Evans}, {Eyer}, {Jansen}, {Jordi}, {Klioner}, {Lammers}, {Lindegren},
  {Luri}, {Mignard}, {Panem}, {Pourbaix}, {Randich}, {Sartoretti}, {Siddiqui},
  {Soubiran}, {van Leeuwen}, {Walton}, {Arenou}, {Bastian}, {Cropper},
  {Drimmel}, {Katz}, {Lattanzi}, {Bakker}, {Cacciari}, {Casta{\~n}eda},
  {Chaoul}, {Cheek}, {De Angeli}, {Fabricius}, {Guerra}, {Holl}, {Masana},
  {Messineo}, {Mowlavi}, {Nienartowicz}, {Panuzzo}, {Portell}, {Riello},
  {Seabroke}, {Tanga}, {Th{\'e}venin}, {Gracia-Abril}, {Comoretto},
  {Garcia-Reinaldos}, {Teyssier}, {Altmann}, {Andrae}, {Audard},
  {Bellas-Velidis}, {Benson}, {Berthier}, {Blomme}, {Burgess}, {Busso},
  {Carry}, {Cellino}, {Clementini}, {Clotet}, {Creevey}, {Davidson}, {De
  Ridder}, {Delchambre}, {Dell'Oro}, {Ducourant},
  {Fern{\'a}ndez-Hern{\'a}ndez}, {Fouesneau}, {Fr{\'e}mat}, {Galluccio},
  {Garc{\'\i}a-Torres}, {Gonz{\'a}lez-N{\'u}{\~n}ez}, {Gonz{\'a}lez-Vidal},
  {Gosset}, {Guy}, {Halbwachs}, {Hambly}, {Harrison}, {Hern{\'a}ndez},
  {Hestroffer}, {Hodgkin}, {Hutton}, {Jasniewicz}, {Jean-Antoine-Piccolo},
  {Jordan}, {Korn}, {Krone-Martins}, {Lanzafame}, {Lebzelter}, {L{\"o}ffler},
  {Manteiga}, {Marrese}, {Mart{\'\i}n-Fleitas}, {Moitinho}, {Mora}, {Muinonen},
  {Osinde}, {Pancino}, {Pauwels}, {Petit}, {Recio-Blanco}, {Richards},
  {Rimoldini}, {Robin}, {Sarro}, {Siopis}, {Smith}, {Sozzetti}, {S{\"u}veges},
  {Torra}, {van Reeven}, {Abbas}, {Abreu Aramburu}, {Accart}, {Aerts},
  {Altavilla}, {{\'A}lvarez}, {Alvarez}, {Alves}, {Anderson}, {Andrei},
  {Anglada Varela}, {Antiche}, {Antoja}, {Arcay}, {Astraatmadja}, {Bach},
  {Baker}, {Balaguer-N{\'u}{\~n}ez}, {Balm}, {Barache}, {Barata}, {Barbato},
  {Barblan}, {Barklem}, {Barrado}, {Barros}, {Barstow}, {Bartholom{\'e}
  Mu{\~n}oz}, {Bassilana}, {Becciani}, {Bellazzini}, {Berihuete}, {Bertone},
  {Bianchi}, {Bienaym{\'e}}, {Blanco-Cuaresma}, {Boch}, {Boeche}, {Bombrun},
  {Borrachero}, {Bossini}, {Bouquillon}, {Bourda}, {Bragaglia}, {Bramante},
  {Breddels}, {Bressan}, {Brouillet}, {Br{\"u}semeister}, {Brugaletta},
  {Bucciarelli}, {Burlacu}, {Busonero}, {Butkevich}, {Buzzi}, {Caffau},
  {Cancelliere}, {Cannizzaro}, {Cantat-Gaudin}, {Carballo}, {Carlucci},
  {Carrasco}, {Casamiquela}, {Castellani}, {Castro-Ginard}, {Charlot},
  {Chemin}, {Chiavassa}, {Cocozza}, {Costigan}, {Cowell}, {Crifo}, {Crosta},
  {Crowley}, {Cuypers}, {Dafonte}, {Damerdji}, {Dapergolas}, {David}, {David},
  {de Laverny}, {De Luise}, {De March}, {de Martino}, {de Souza}, {de Torres},
  {Debosscher}, {del Pozo}, {Delbo}, {Delgado}, {Delgado}, {Di Matteo},
  {Diakite}, {Diener}, {Distefano}, {Dolding}, {Drazinos}, {Dur{\'a}n},
  {Edvardsson}, {Enke}, {Eriksson}, {Esquej}, {Eynard Bontemps}, {Fabre},
  {Fabrizio}, {Faigler}, {Falc{\~a}o}, {Farr{\`a}s Casas}, {Federici},
  {Fedorets}, {Fernique}, {Figueras}, {Filippi}, {Findeisen}, {Fonti},
  {Fraile}, {Fraser}, {Fr{\'e}zouls}, {Gai}, {Galleti}, {Garabato},
  {Garc{\'\i}a-Sedano}, {Garofalo}, {Garralda}, {Gavel}, {Gavras}, {Gerssen},
  {Geyer}, {Giacobbe}, {Gilmore}, {Girona}, {Giuffrida}, {Glass}, {Gomes},
  {Granvik}, {Gueguen}, {Guerrier}, {Guiraud}, {Guti{\'e}rrez-S{\'a}nchez},
  {Haigron}, {Hatzidimitriou}, {Hauser}, {Haywood}, {Heiter}, {Helmi}, {Heu},
  {Hilger}, {Hobbs}, {Hofmann}, {Holland}, {Huckle}, {Hypki}, {Icardi},
  {Jan{\ss}en}, {Jevardat de Fombelle}, {Jonker}, {Juh{\'a}sz}, {Julbe},
  {Karampelas}, {Kewley}, {Klar}, {Kochoska}, {Kohley}, {Kolenberg},
  {Kontizas}, {Kontizas}, {Koposov}, {Kordopatis}, {Kostrzewa-Rutkowska},
  {Koubsky}, {Lambert}, {Lanza}, {Lasne}, {Lavigne}, {Le Fustec}, {Le
  Poncin-Lafitte}, {Lebreton}, {Leccia}, {Leclerc}, {Lecoeur-Taibi},
  {Lenhardt}, {Leroux}, {Liao}, {Licata}, {Lindstr{\o}m}, {Lister}, {Livanou},
  {Lobel}, {L{\'o}pez}, {Managau}, {Mann}, {Mantelet}, {Marchal}, {Marchant},
  {Marconi}, {Marinoni}, {Marschalk{\'o}}, {Marshall}, {Martino}, {Marton},
  {Mary}, {Massari}, {Matijevi{\v{c}}}, {Mazeh}, {McMillan}, {Messina},
  {Michalik}, {Millar}, {Molina}, {Molinaro}, {Moln{\'a}r}, {Montegriffo},
  {Mor}, {Morbidelli}, {Morel}, {Morris}, {Mulone}, {Muraveva}, {Musella},
  {Nelemans}, {Nicastro}, {Noval}, {O'Mullane}, {Ord{\'e}novic},
  {Ord{\'o}{\~n}ez-Blanco}, {Osborne}, {Pagani}, {Pagano}, {Pailler},
  {Palacin}, {Palaversa}, {Panahi}, {Pawlak}, {Piersimoni}, {Pineau}, {Plachy},
  {Plum}, {Poggio}, {Poujoulet}, {Pr{\v{s}}a}, {Pulone}, {Racero}, {Ragaini},
  {Rambaux}, {Ramos-Lerate}, {Regibo}, {Reyl{\'e}}, {Riclet}, {Ripepi}, {Riva},
  {Rivard}, {Rixon}, {Roegiers}, {Roelens}, {Romero-G{\'o}mez}, {Rowell},
  {Royer}, {Ruiz-Dern}, {Sadowski}, {Sagrist{\`a} Sell{\'e}s}, {Sahlmann},
  {Salgado}, {Salguero}, {Sanna}, {Santana-Ros}, {Sarasso}, {Savietto},
  {Schultheis}, {Sciacca}, {Segol}, {Segovia}, {S{\'e}gransan}, {Shih},
  {Siltala}, {Silva}, {Smart}, {Smith}, {Solano}, {Solitro}, {Sordo}, {Soria
  Nieto}, {Souchay}, {Spagna}, {Spoto}, {Stampa}, {Steele},
  {Steidelm{\"u}ller}, {Stephenson}, {Stoev}, {Suess}, {Surdej}, {Szabados},
  {Szegedi-Elek}, {Tapiador}, {Taris}, {Tauran}, {Taylor}, {Teixeira},
  {Terrett}, {Teyssand ier}, {Thuillot}, {Titarenko}, {Torra Clotet}, {Turon},
  {Ulla}, {Utrilla}, {Uzzi}, {Vaillant}, {Valentini}, {Valette}, {van Elteren},
  {Van Hemelryck}, {van Leeuwen}, {Vaschetto}, {Vecchiato}, {Veljanoski},
  {Viala}, {Vicente}, {Vogt}, {von Essen}, {Voss}, {Votruba}, {Voutsinas},
  {Walmsley}, {Weiler}, {Wertz}, {Wevers}, {Wyrzykowski}, {Yoldas},
  {{\v{Z}}erjal}, {Ziaeepour}, {Zorec}, {Zschocke}, {Zucker}, {Zurbach}, \&
  {Zwitter}}]{gaia2}
{Gaia Collaboration}, {Brown}, A.~G.~A., {Vallenari}, A., {et~al.} 2018, \aap,
  616, A1

\bibitem[{{Garc{\'\i}a P{\'e}rez} {et~al.}(2016){Garc{\'\i}a P{\'e}rez},
  {Allende Prieto}, {Holtzman}, {Shetrone}, {M{\'e}sz{\'a}ros}, {Bizyaev},
  {Carrera}, {Cunha}, {Garc{\'\i}a-Hern{\'a}ndez}, {Johnson}, {Majewski},
  {Nidever}, {Schiavon}, {Shane}, {Smith}, {Sobeck}, {Troup}, {Zamora},
  {Weinberg}, {Bovy}, {Eisenstein}, {Feuillet}, {Frinchaboy}, {Hayden},
  {Hearty}, {Nguyen}, {O'Connell}, {Pinsonneault}, {Wilson}, \&
  {Zasowski}}]{aspcap}
{Garc{\'\i}a P{\'e}rez}, A.~E., {Allende Prieto}, C., {Holtzman}, J.~A.,
  {et~al.} 2016, \aj, 151, 144

\bibitem[{{Gilmore} \& {Reid}(1983)}]{GilmoreReid1983}
{Gilmore}, G., \& {Reid}, N. 1983, \mnras, 202, 1025

\bibitem[{{Guggenberger} {et~al.}(2017){Guggenberger}, {Hekker}, {Angelou},
  {Basu}, \& {Bellinger}}]{guggenberg+2017}
{Guggenberger}, E., {Hekker}, S., {Angelou}, G.~C., {Basu}, S., \& {Bellinger},
  E.~P. 2017, \mnras, 470, 2069

\bibitem[{{Gunn} {et~al.}(2006){Gunn}, {Siegmund}, {Mannery}, {Owen}, {Hull},
  {Leger}, {Carey}, {Knapp}, {York}, {Boroski}, {Kent}, {Lupton}, {Rockosi},
  {Evans}, {Waddell}, {Anderson}, {Annis}, {Barentine}, {Bartoszek}, {Bastian},
  {Bracker}, {Brewington}, {Briegel}, {Brinkmann}, {Brown}, {Carr},
  {Czarapata}, {Drennan}, {Dombeck}, {Federwitz}, {Gillespie}, {Gonzales},
  {Hansen}, {Harvanek}, {Hayes}, {Jordan}, {Kinney}, {Klaene}, {Kleinman},
  {Kron}, {Kresinski}, {Lee}, {Limmongkol}, {Lindenmeyer}, {Long}, {Loomis},
  {McGehee}, {Mantsch}, {Neilsen}, {Neswold}, {Newman}, {Nitta}, {Peoples},
  {Pier}, {Prieto}, {Prosapio}, {Rivetta}, {Schneider}, {Snedden}, \&
  {Wang}}]{gunn2006}
{Gunn}, J.~E., {Siegmund}, W.~A., {Mannery}, E.~J., {et~al.} 2006, \aj, 131,
  2332

\bibitem[{{Hayden} {et~al.}(2017){Hayden}, {Recio-Blanco}, {de Laverny},
  {Mikolaitis}, \& {Worley}}]{haydenambre}
{Hayden}, M.~R., {Recio-Blanco}, A., {de Laverny}, P., {Mikolaitis}, S., \&
  {Worley}, C.~C. 2017, \aap, 608, L1

\bibitem[{{Hayden} {et~al.}(2015){Hayden}, {Bovy}, {Holtzman}, {Nidever},
  {Bird}, {Weinberg}, {Andrews}, {Majewski}, {Allende Prieto}, {Anders},
  {Beers}, {Bizyaev}, {Chiappini}, {Cunha}, {Frinchaboy},
  {Garc{\'\i}a-Her{\'n}and ez}, {Garc{\'\i}a P{\'e}rez}, {Girardi}, {Harding},
  {Hearty}, {Johnson}, {M{\'e}sz{\'a}ros}, {Minchev}, {O'Connell}, {Pan},
  {Robin}, {Schiavon}, {Schneider}, {Schultheis}, {Shetrone}, {Skrutskie},
  {Steinmetz}, {Smith}, {Wilson}, {Zamora}, \& {Zasowski}}]{Hayden2015}
{Hayden}, M.~R., {Bovy}, J., {Holtzman}, J.~A., {et~al.} 2015, \apj, 808, 132

\bibitem[{{Hekker} {et~al.}(2010){Hekker}, {Broomhall}, {Chaplin}, {Elsworth},
  {Fletcher}, {New}, {Arentoft}, {Quirion}, \& {Kjeldsen}}]{BHM}
{Hekker}, S., {Broomhall}, A.~M., {Chaplin}, W.~J., {et~al.} 2010, \mnras, 402,
  2049

\bibitem[{{Holtzman} {et~al.}(2018){Holtzman}, {Hasselquist}, {Shetrone},
  {Cunha}, {Allende Prieto}, {Anguiano}, {Bizyaev}, {Bovy}, {Casey},
  {Edvardsson}, {Johnson}, {J{\"o}nsson}, {Meszaros}, {Smith}, {Sobeck},
  {Zamora}, {Chojnowski}, {Fernandez-Trincado}, {Garcia-Hernandez}, {Majewski},
  {Pinsonneault}, {Souto}, {Stringfellow}, {Tayar}, {Troup}, \&
  {Zasowski}}]{holtzman2018}
{Holtzman}, J.~A., {Hasselquist}, S., {Shetrone}, M., {et~al.} 2018, \aj, 156,
  125

\bibitem[{{Howell} {et~al.}(2014){Howell}, {Sobeck}, {Haas}, {Still},
  {Barclay}, {Mullally}, {Troeltzsch}, {Aigrain}, {Bryson}, {Caldwell},
  {Chaplin}, {Cochran}, {Huber}, {Marcy}, {Miglio}, {Najita}, {Smith},
  {Twicken}, \& {Fortney}}]{k2mission}
{Howell}, S.~B., {Sobeck}, C., {Haas}, M., {et~al.} 2014, \pasp, 126, 398

\bibitem[{{Huber} {et~al.}(2009){Huber}, {Stello}, {Bedding}, {Chaplin},
  {Arentoft}, {Quirion}, \& {Kjeldsen}}]{SYD}
{Huber}, D., {Stello}, D., {Bedding}, T.~R., {et~al.} 2009, Communications in
  Asteroseismology, 160, 74

\bibitem[{{Huber} {et~al.}(2017){Huber}, {Zinn}, {Bojsen-Hansen},
  {Pinsonneault}, {Sahlholdt}, {Serenelli}, {Silva Aguirre}, {Stassun},
  {Stello}, {Tayar}, {Bastien}, {Bedding}, {Buchhave}, {Chaplin}, {Davies},
  {Garc{\'{\i}}a}, {Latham}, {Mathur}, {Mosser}, \& {Sharma}}]{huber+2017}
{Huber}, D., {Zinn}, J., {Bojsen-Hansen}, M., {et~al.} 2017, \apj, 844, 102

\bibitem[{{J{\"o}nsson} {et~al.}(2020){J{\"o}nsson}, {Holtzman}, {Allende
  Prieto}, {Cunha}, {Garc{\'\i}a-Hern{\'a}ndez}, {Hasselquist}, {Masseron},
  {Osorio}, {Shetrone}, {Smith}, {Stringfellow}, {Bizyaev}, {Edvardsson},
  {Majewski}, {M{\'e}sz{\'a}ros}, {Souto}, {Zamora}, {Beaton}, {Bovy}, {Donor},
  {Pinsonneault}, {Poovelil}, \& {Sobeck}}]{jonsson2020}
{J{\"o}nsson}, H., {Holtzman}, J.~A., {Allende Prieto}, C., {et~al.} 2020, \aj,
  160, 120

\bibitem[{{Kallinger} {et~al.}(2016){Kallinger}, {Hekker}, {Garcia}, {Huber},
  \& {Matthews}}]{CAN}
{Kallinger}, T., {Hekker}, S., {Garcia}, R.~A., {Huber}, D., \& {Matthews},
  J.~M. 2016, Science Advances, 2, 1500654

\bibitem[{{Khan} {et~al.}(2019){Khan}, {Miglio}, {Mosser}, {Arenou},
  {Belkacem}, {Brown}, {Katz}, {Casagrand e}, {Chaplin}, {Davies}, {Rendle},
  {Rodrigues}, {Bossini}, {Cantat-Gaudin}, {Elsworth}, {Girardi}, {North}, \&
  {Vallenari}}]{Khan2019}
{Khan}, S., {Miglio}, A., {Mosser}, B., {et~al.} 2019, \aap, 628, A35

\bibitem[{{Kjeldsen} \& {Bedding}(1995)}]{kjeldsen&bedding1995}
{Kjeldsen}, H., \& {Bedding}, T.~R. 1995, \aap, 293, 87

\bibitem[{Lian {et~al.}(2020)Lian, Thomas, Maraston, Zamora, Tayar, Pan,
  Tissera, Fern{\'{a}}ndez-Trincado, \& Garcia-Hernandez}]{lian+2020}
Lian, J., Thomas, D., Maraston, C., {et~al.} 2020, \mnras, 494, 2561

\bibitem[{{Majewski} {et~al.}(2017){Majewski}, {Schiavon}, {Frinchaboy},
  {Allende Prieto}, {Barkhouser}, {Bizyaev}, {Blank}, {Brunner}, {Burton},
  {Carrera}, {Chojnowski}, {Cunha}, {Epstein}, {Fitzgerald}, {Garc{\'\i}a
  P{\'e}rez}, {Hearty}, {Henderson}, {Holtzman}, {Johnson}, {Lam}, {Lawler},
  {Maseman}, {M{\'e}sz{\'a}ros}, {Nelson}, {Nguyen}, {Nidever}, {Pinsonneault},
  {Shetrone}, {Smee}, {Smith}, {Stolberg}, {Skrutskie}, {Walker}, {Wilson},
  {Zasowski}, {Anders}, {Basu}, {Beland}, {Blanton}, {Bovy}, {Brownstein},
  {Carlberg}, {Chaplin}, {Chiappini}, {Eisenstein}, {Elsworth}, {Feuillet},
  {Fleming}, {Galbraith-Frew}, {Garc{\'\i}a}, {Garc{\'\i}a-Hern{\'a}ndez},
  {Gillespie}, {Girardi}, {Gunn}, {Hasselquist}, {Hayden}, {Hekker}, {Ivans},
  {Kinemuchi}, {Klaene}, {Mahadevan}, {Mathur}, {Mosser}, {Muna}, {Munn},
  {Nichol}, {O'Connell}, {Parejko}, {Robin}, {Rocha-Pinto}, {Schultheis},
  {Serenelli}, {Shane}, {Silva Aguirre}, {Sobeck}, {Thompson}, {Troup},
  {Weinberg}, \& {Zamora}}]{majewski2017}
{Majewski}, S.~R., {Schiavon}, R.~P., {Frinchaboy}, P.~M., {et~al.} 2017, \aj,
  154, 94

\bibitem[{{Mamajek} {et~al.}(2015){Mamajek}, {Prsa}, {Torres}, {Harmanec},
  {Asplund}, {Bennett}, {Capitaine}, {Christensen-Dalsgaard}, {Depagne},
  {Folkner}, {Haberreiter}, {Hekker}, {Hilton}, {Kostov}, {Kurtz}, {Laskar},
  {Mason}, {Milone}, {Montgomery}, {Richards}, {Schou}, \&
  {Stewart}}]{mamajek+2015}
{Mamajek}, E.~E., {Prsa}, A., {Torres}, G., {et~al.} 2015, arXiv e-prints

\bibitem[{{Martig} {et~al.}(2015){Martig}, {Rix}, {Silva Aguirre}, {Hekker},
  {Mosser}, {Elsworth}, {Bovy}, {Stello}, {Anders}, {Garc{\'\i}a}, {Tayar},
  {Rodrigues}, {Basu}, {Carrera}, {Ceillier}, {Chaplin}, {Chiappini},
  {Frinchaboy}, {Garc{\'\i}a-Hern{\'a}ndez}, {Hearty}, {Holtzman}, {Johnson},
  {Majewski}, {Mathur}, {M{\'e}sz{\'a}ros}, {Miglio}, {Nidever}, {Pan},
  {Pinsonneault}, {Schiavon}, {Schneider}, {Serenelli}, {Shetrone}, \&
  {Zamora}}]{Martig2015}
{Martig}, M., {Rix}, H.-W., {Silva Aguirre}, V., {et~al.} 2015, \mnras, 451,
  2230

\bibitem[{{Mathur} {et~al.}(2010){Mathur}, {Garc{\'\i}a}, {R{\'e}gulo},
  {Creevey}, {Ballot}, {Salabert}, {Arentoft}, {Quirion}, {Chaplin}, \&
  {Kjeldsen}}]{A2Z}
{Mathur}, S., {Garc{\'\i}a}, R.~A., {R{\'e}gulo}, C., {et~al.} 2010, \aap, 511,
  A46

\bibitem[{{Miglio} {et~al.}(2020){Miglio}, {Chiappini}, {Mackereth}, {Davies},
  {Brogaard}, {Casagrande}, {Chaplin}, {Girardi}, {Kawata}, {Khan}, {Izzard},
  {Montalban}, {Mosser}, {Vincenzo}, {Bossini}, {Noels}, {Rodrigues},
  {Valentini}, \& {Mand el}}]{miglio2020}
{Miglio}, A., {Chiappini}, C., {Mackereth}, T., {et~al.} 2020, arXiv e-prints

\bibitem[{{Montalb{\'a}n} {et~al.}(2020){Montalb{\'a}n}, {Mackereth}, {Miglio},
  {Vincenzo}, {Chiappini}, {Buldgen}, {Mosser}, {Noels}, {Scuflaire}, {Vrard},
  {Willett}, {Davies}, {Hall}, {Nielsen}, {Khan}, {Rendle}, {van Rossem},
  {Ferguson}, \& {Chaplin}}]{montalban+2020}
{Montalb{\'a}n}, J., {Mackereth}, J.~T., {Miglio}, A., {et~al.} 2020, arXiv
  e-prints, arXiv:2006.01783

\bibitem[{{Mosser} \& {Appourchaux}(2009)}]{COR}
{Mosser}, B., \& {Appourchaux}, T. 2009, \aap, 508, 877

\bibitem[{{Mosser} {et~al.}(2013){Mosser}, {Dziembowski}, {Belkacem}, {Goupil},
  {Michel}, {Samadi}, {Soszy{\'n}ski}, {Vrard}, {Elsworth}, {Hekker}, \&
  {Mathur}}]{Mosser2013}
{Mosser}, B., {Dziembowski}, W.~A., {Belkacem}, K., {et~al.} 2013, \aap, 559,
  A137

\bibitem[{{Nidever} {et~al.}(2014){Nidever}, {Bovy}, {Bird}, {Andrews},
  {Hayden}, {Holtzman}, {Majewski}, {Smith}, {Robin}, {Garc{\'\i}a P{\'e}rez},
  {Cunha}, {Allende Prieto}, {Zasowski}, {Schiavon}, {Johnson}, {Weinberg},
  {Feuillet}, {Schneider}, {Shetrone}, {Sobeck}, {Garc{\'\i}a-Hern{\'a}ndez},
  {Zamora}, {Rix}, {Beers}, {Wilson}, {O'Connell}, {Minchev}, {Chiappini},
  {Anders}, {Bizyaev}, {Brewington}, {Ebelke}, {Frinchaboy}, {Ge}, {Kinemuchi},
  {Malanushenko}, {Malanushenko}, {Marchante}, {M{\'e}sz{\'a}ros}, {Oravetz},
  {Pan}, {Simmons}, \& {Skrutskie}}]{NBB2014}
{Nidever}, D.~L., {Bovy}, J., {Bird}, J.~C., {et~al.} 2014, \apj, 796, 38

\bibitem[{{Nidever} {et~al.}(2015){Nidever}, {Holtzman}, {Allende Prieto},
  {Beland}, {Bender}, {Bizyaev}, {Burton}, {Desphande}, {Fleming}, {Garc{\'\i}a
  P{\'e}rez}, {Hearty}, {Majewski}, {M{\'e}sz{\'a}ros}, {Muna}, {Nguyen},
  {Schiavon}, {Shetrone}, {Skrutskie}, {Sobeck}, \& {Wilson}}]{Nidever2015}
{Nidever}, D.~L., {Holtzman}, J.~A., {Allende Prieto}, C., {et~al.} 2015, \aj,
  150, 173

\bibitem[{{Pinsonneault} {et~al.}(1989){Pinsonneault}, {Kawaler}, {Sofia}, \&
  {Demarque}}]{yrec}
{Pinsonneault}, M.~H., {Kawaler}, S.~D., {Sofia}, S., \& {Demarque}, P. 1989,
  \apj, 338, 424

\bibitem[{{Pinsonneault} {et~al.}(2014){Pinsonneault}, {Elsworth}, {Epstein},
  {Hekker}, {M{\'e}sz{\'a}ros}, {Chaplin}, {Johnson}, {Garc{\'\i}a},
  {Holtzman}, {Mathur}, {Garc{\'\i}a P{\'e}rez}, {Silva Aguirre}, {Girardi},
  {Basu}, {Shetrone}, {Stello}, {Allende Prieto}, {An}, {Beck}, {Beers},
  {Bizyaev}, {Bloemen}, {Bovy}, {Cunha}, {De Ridder}, {Frinchaboy},
  {Garc{\'\i}a-Hern{\'a}ndez}, {Gilliland}, {Harding}, {Hearty}, {Huber},
  {Ivans}, {Kallinger}, {Majewski}, {Metcalfe}, {Miglio}, {Mosser}, {Muna},
  {Nidever}, {Schneider}, {Serenelli}, {Smith}, {Tayar}, {Zamora}, \&
  {Zasowski}}]{apokasc1}
{Pinsonneault}, M.~H., {Elsworth}, Y., {Epstein}, C., {et~al.} 2014, \apjs,
  215, 19

\bibitem[{{Pinsonneault} {et~al.}(2018){Pinsonneault}, {Elsworth}, {Tayar},
  {Serenelli}, {Stello}, {Zinn}, {Mathur}, {Garc{\'\i}a}, {Johnson}, {Hekker},
  {Huber}, {Kallinger}, {M{\'e}sz{\'a}ros}, {Mosser}, {Stassun}, {Girardi},
  {Rodrigues}, {Silva Aguirre}, {An}, {Basu}, {Chaplin}, {Corsaro}, {Cunha},
  {Garc{\'\i}a-Hern{\'a}ndez}, {Holtzman}, {J{\"o}nsson}, {Shetrone}, {Smith},
  {Sobeck}, {Stringfellow}, {Zamora}, {Beers}, {Fern{\'a}ndez-Trincado},
  {Frinchaboy}, {Hearty}, \& {Nitschelm}}]{apokasc2}
{Pinsonneault}, M.~H., {Elsworth}, Y.~P., {Tayar}, J., {et~al.} 2018, \apjs,
  239, 32

\bibitem[{{Price-Whelan} {et~al.}(2018){Price-Whelan}, {Sip{\H{o}}cz},
  {G{\"u}nther}, {Lim}, {Crawford}, {Conseil}, {Shupe}, {Craig}, {Dencheva},
  {Ginsburg}, {VanderPlas}, {Bradley}, {P{\'e}rez-Su{\'a}rez}, {de Val-Borro},
  {Paper Contributors}, {Aldcroft}, {Cruz}, {Robitaille}, {Tollerud},
  {Coordination Committee}, {Ardelean}, {Babej}, {Bach}, {Bachetti}, {Bakanov},
  {Bamford}, {Barentsen}, {Barmby}, {Baumbach}, {Berry}, {Biscani}, {Boquien},
  {Bostroem}, {Bouma}, {Brammer}, {Bray}, {Breytenbach}, {Buddelmeijer},
  {Burke}, {Calderone}, {Cano Rodr{\'\i}guez}, {Cara}, {Cardoso}, {Cheedella},
  {Copin}, {Corrales}, {Crichton}, {D{\textquoteright}Avella}, {Deil},
  {Depagne}, {Dietrich}, {Donath}, {Droettboom}, {Earl}, {Erben}, {Fabbro},
  {Ferreira}, {Finethy}, {Fox}, {Garrison}, {Gibbons}, {Goldstein}, {Gommers},
  {Greco}, {Greenfield}, {Groener}, {Grollier}, {Hagen}, {Hirst}, {Homeier},
  {Horton}, {Hosseinzadeh}, {Hu}, {Hunkeler}, {Ivezi{\'c}}, {Jain}, {Jenness},
  {Kanarek}, {Kendrew}, {Kern}, {Kerzendorf}, {Khvalko}, {King}, {Kirkby},
  {Kulkarni}, {Kumar}, {Lee}, {Lenz}, {Littlefair}, {Ma}, {Macleod},
  {Mastropietro}, {McCully}, {Montagnac}, {Morris}, {Mueller}, {Mumford},
  {Muna}, {Murphy}, {Nelson}, {Nguyen}, {Ninan}, {N{\"o}the}, {Ogaz}, {Oh},
  {Parejko}, {Parley}, {Pascual}, {Patil}, {Patil}, {Plunkett}, {Prochaska},
  {Rastogi}, {Reddy Janga}, {Sabater}, {Sakurikar}, {Seifert}, {Sherbert},
  {Sherwood-Taylor}, {Shih}, {Sick}, {Silbiger}, {Singanamalla}, {Singer},
  {Sladen}, {Sooley}, {Sornarajah}, {Streicher}, {Teuben}, {Thomas},
  {Tremblay}, {Turner}, {Terr{\'o}n}, {van Kerkwijk}, {de la Vega}, {Watkins},
  {Weaver}, {Whitmore}, {Woillez}, {Zabalza}, \& {Contributors}}]{astropy:2018}
{Price-Whelan}, A.~M., {Sip{\H{o}}cz}, B.~M., {G{\"u}nther}, H.~M., {et~al.}
  2018, \aj, 156, 123

\bibitem[{{Prochaska} {et~al.}(2000){Prochaska}, {Naumov}, {Carney},
  {McWilliam}, \& {Wolfe}}]{prochaska2000}
{Prochaska}, J.~X., {Naumov}, S.~O., {Carney}, B.~W., {McWilliam}, A., \&
  {Wolfe}, A.~M. 2000, \aj, 120, 2513

\bibitem[{{Rendle} {et~al.}(2019){Rendle}, {Miglio}, {Chiappini}, {Valentini},
  {Davies}, {Mosser}, {Elsworth}, {Garc{\'\i}a}, {Mathur}, {Jofr{\'e}},
  {Worley}, {Casagrande}, {Girardi}, {Lund}, {Feuillet}, {Gavel}, {Magrini},
  {Khan}, {Rodrigues}, {Johnson}, {Cunha}, {Lane}, {Nitschelm}, \&
  {Chaplin}}]{rendle+2019}
{Rendle}, B.~M., {Miglio}, A., {Chiappini}, C., {et~al.} 2019, \mnras, 490,
  4465

\bibitem[{{Rodrigues} {et~al.}(2014){Rodrigues}, {Girardi}, {Miglio},
  {Bossini}, {Bovy}, {Epstein}, {Pinsonneault}, {Stello}, {Zasowski}, {Allende
  Prieto}, {Chaplin}, {Hekker}, {Johnson}, {M{\'e}sz{\'a}ros}, {Mosser},
  {Anders}, {Basu}, {Beers}, {Chiappini}, {da Costa}, {Elsworth},
  {Garc{\'\i}a}, {Garc{\'\i}a P{\'e}rez}, {Hearty}, {Maia}, {Majewski},
  {Mathur}, {Montalb{\'a}n}, {Nidever}, {Santiago}, {Schultheis}, {Serenelli},
  \& {Shetrone}}]{param2}
{Rodrigues}, T.~S., {Girardi}, L., {Miglio}, A., {et~al.} 2014, \mnras, 445,
  2758

\bibitem[{{Salaris} {et~al.}(1993){Salaris}, {Chieffi}, \&
  {Straniero}}]{salaris+1993}
{Salaris}, M., {Chieffi}, A., \& {Straniero}, O. 1993, \apj, 414, 580

\bibitem[{{Sch{\"o}nrich} \& {Binney}(2009)}]{SB2009}
{Sch{\"o}nrich}, R., \& {Binney}, J. 2009, \mnras, 396, 203

\bibitem[{{Sellwood} \& {Binney}(2002)}]{sellwoodbinney2002}
{Sellwood}, J.~A., \& {Binney}, J.~J. 2002, \mnras, 336, 785

\bibitem[{{Sharma} {et~al.}(2020){Sharma}, {Hayden}, \&
  {Bland-Hawthorn}}]{sharma+2020b}
{Sharma}, S., {Hayden}, M.~R., \& {Bland-Hawthorn}, J. 2020, arXiv e-prints,
  arXiv:2005.03646

\bibitem[{{Sharma} {et~al.}(2016){Sharma}, {Stello}, {Bland-Hawthorn}, {Huber
  }, \& {Bedding}}]{sharma+2016}
{Sharma}, S., {Stello}, D., {Bland-Hawthorn}, J., {Huber }, D., \& {Bedding},
  T.~R. 2016, \apj, 822, 15

\bibitem[{{Silva Aguirre} {et~al.}(2015){Silva Aguirre}, {Davies}, {Basu},
  {Christensen-Dalsgaard}, {Creevey}, {Metcalfe}, {Bedding}, {Casagrande},
  {Handberg}, {Lund}, {Nissen}, {Chaplin}, {Huber}, {Serenelli}, {Stello}, {Van
  Eylen}, {Campante}, {Elsworth}, {Gilliland}, {Hekker}, {Karoff}, {Kawaler},
  {Kjeldsen}, \& {Lundkvist}}]{silva_aguirre+2015}
{Silva Aguirre}, V., {Davies}, G.~R., {Basu}, S., {et~al.} 2015, \mnras, 452,
  2127

\bibitem[{{Silva Aguirre} {et~al.}(2017){Silva Aguirre}, {Lund}, {Antia},
  {Ball}, {Basu}, {Christensen-Dalsgaard}, {Lebreton}, {Reese}, {Verma},
  {Casagrande}, {Justesen}, {Mosumgaard}, {Chaplin}, {Bedding}, {Davies},
  {Handberg}, {Houdek}, {Huber}, {Kjeldsen}, {Latham}, {White}, {Coelho},
  {Miglio}, \& {Rendle}}]{silva_aguirre+2017}
{Silva Aguirre}, V., {Lund}, M.~N., {Antia}, H.~M., {et~al.} 2017, \apj, 835,
  173

\bibitem[{{Silva Aguirre} {et~al.}(2018){Silva Aguirre}, {Bojsen-Hansen},
  {Slumstrup}, {Casagrande}, {Kawata}, {Ciuc{\v{a}}}, {Hand berg}, {Lund},
  {Mosumgaard}, {Huber}, {Johnson}, {Pinsonneault}, {Serenelli}, {Stello},
  {Tayar}, {Bird}, {Cassisi}, {Hon}, {Martig}, {Nissen}, {Rix},
  {Sch{\"o}nrich}, {Sahlholdt}, {Trick}, \& {Yu}}]{SilvaAguirre2018}
{Silva Aguirre}, V., {Bojsen-Hansen}, M., {Slumstrup}, D., {et~al.} 2018,
  \mnras, 475, 5487

\bibitem[{{Spitoni} {et~al.}(2019){Spitoni}, {Silva Aguirre}, {Matteucci},
  {Calura}, \& {Grisoni}}]{Spitoni2019}
{Spitoni}, E., {Silva Aguirre}, V., {Matteucci}, F., {Calura}, F., \&
  {Grisoni}, V. 2019, \aap, 623, A60

\bibitem[{{Stello} {et~al.}(2015){Stello}, {Huber}, {Sharma}, {Johnson},
  {Lund}, {Handberg}, {Buzasi}, {Silva Aguirre}, {Chaplin}, {Miglio},
  {Pinsonneault}, {Basu}, {Bedding}, {Bland-Hawthorn}, {Casagrande}, {Davies},
  {Elsworth}, {Garcia}, {Mathur}, {Di Mauro}, {Mosser}, {Schneider},
  {Serenelli}, \& {Valentini}}]{stello+2015}
{Stello}, D., {Huber}, D., {Sharma}, S., {et~al.} 2015, \apjl, 809, L3

\bibitem[{{Stello} {et~al.}(2017){Stello}, {Zinn}, {Elsworth}, {Garcia},
  {Kallinger}, {Mathur}, {Mosser}, {Sharma}, {Chaplin}, {Davies}, {Huber},
  {Jones}, {Miglio}, \& {Silva Aguirre}}]{stello+2017}
{Stello}, D., {Zinn}, J., {Elsworth}, Y., {et~al.} 2017, \apj, 835, 83

\bibitem[{{Tayar} {et~al.}(2017){Tayar}, {Somers}, {Pinsonneault}, {Stello},
  {Mints}, {Johnson}, {Zamora}, {Garc{\'\i}a-Hern{\'a}ndez}, {Maraston},
  {Serenelli}, {Allende Prieto}, {Bastien}, {Basu}, {Bird}, {Cohen}, {Cunha},
  {Elsworth}, {Garc{\'\i}a}, {Girardi}, {Hekker}, {Holtzman}, {Huber},
  {Mathur}, {M{\'e}sz{\'a}ros}, {Mosser}, {Shetrone}, {Silva Aguirre},
  {Stassun}, {Stringfellow}, {Zasowski}, \& {Roman-Lopes}}]{Tayar2017}
{Tayar}, J., {Somers}, G., {Pinsonneault}, M.~H., {et~al.} 2017, \apj, 840, 17

\bibitem[{{Tinsley}(1979)}]{Tinsley1979}
{Tinsley}, B.~M. 1979, \apj, 229, 1046

\bibitem[{{Ulrich}(1986)}]{ulrich1986}
{Ulrich}, R.~K. 1986, \apjl, 306, L37

\bibitem[{{Valentini} {et~al.}(2019){Valentini}, {Chiappini}, {Bossini},
  {Miglio}, {Davies}, {Mosser}, {Elsworth}, {Mathur}, {Garc{\'\i}a}, {Girardi},
  {Rodrigues}, {Steinmetz}, \& {Vallenari}}]{Valentini2019}
{Valentini}, M., {Chiappini}, C., {Bossini}, D., {et~al.} 2019, \aap, 627, A173

\bibitem[{{van Saders} \& {Pinsonneault}(2012)}]{tracks}
{van Saders}, J.~L., \& {Pinsonneault}, M.~H. 2012, \apj, 746, 16

\bibitem[{{Wallerstein}(1962)}]{wallerstein1962}
{Wallerstein}, G. 1962, \apjs, 6, 407

\bibitem[{Waskom {et~al.}(2017)Waskom, Botvinnik, O'Kane, Hobson, Lukauskas,
  Gemperline, Augspurger, Halchenko, Cole, Warmenhoven, de~Ruiter, Pye, Hoyer,
  Vanderplas, Villalba, Kunter, Quintero, Bachant, Martin, Meyer, Miles, Ram,
  Yarkoni, Williams, Evans, Fitzgerald, Brian, Fonnesbeck, Lee, \&
  Qalieh}]{seaborn}
Waskom, M., Botvinnik, O., O'Kane, D., {et~al.} 2017, mwaskom/seaborn: v0.8.1
  (September 2017), v0.8.1,  Zenodo

\bibitem[{{Weinberg} {et~al.}(2017){Weinberg}, {Andrews}, \&
  {Freudenburg}}]{Weinberg+2017}
{Weinberg}, D.~H., {Andrews}, B.~H., \& {Freudenburg}, J. 2017, \apj, 837, 183

\bibitem[{{Weinberg} {et~al.}(2019){Weinberg}, {Holtzman}, {Hasselquist},
  {Bird}, {Johnson}, {Shetrone}, {Sobeck}, {Allende Prieto}, {Bizyaev},
  {Carrera}, {Cohen}, {Cunha}, {Ebelke}, {Fernandez-Trincado},
  {Garc{\'\i}a-Hern{\'a}ndez}, {Hayes}, {J{\"o}nsson}, {Lane}, {Majewski},
  {Malanushenko}, {M{\'e}sz{\'a}ros}, {Nidever}, {Nitschelm}, {Pan}, {Rix},
  {Rybizki}, {Schiavon}, {Schneider}, {Wilson}, \& {Zamora}}]{Weinberg2019}
{Weinberg}, D.~H., {Holtzman}, J.~A., {Hasselquist}, S., {et~al.} 2019, \apj,
  874, 102

\bibitem[{White {et~al.}(2011)White, Bedding, Stello, Christensen-Dalsgaard,
  Huber, \& Kjeldsen}]{white+2011}
White, T.~R., Bedding, T.~R., Stello, D., {et~al.} 2011, The Astrophysical
  Journal, 743, 161

\bibitem[{{Wilson} {et~al.}(2019){Wilson}, {Hearty}, {Skrutskie}, {Majewski},
  {Holtzman}, {Eisenstein}, {Gunn}, {Blank}, {Henderson}, {Smee}, {Nelson},
  {Nidever}, {Arns}, {Barkhouser}, {Barr}, {Beland}, {Bershady}, {Blanton},
  {Brunner}, {Burton}, {Carey}, {Carr}, {Colque}, {Crane}, {Damke}, {Davidson},
  {Dean}, {Di Mille}, {Don}, {Ebelke}, {Evans}, {Fitzgerald}, {Gillespie},
  {Hall}, {Harding}, {Harding}, {Hammond}, {Hancock}, {Harrison}, {Hope},
  {Horne}, {Karakla}, {Lam}, {Leger}, {MacDonald}, {Maseman}, {Matsunari},
  {Melton}, {Mitcheltree}, {O'Brien}, {O'Connell}, {Patten}, {Richardson},
  {Rieke}, {Rieke}, {Roman-Lopes}, {Schiavon}, {Sobeck}, {Stolberg}, {Stoll},
  {Tembe}, {Trujillo}, {Uomoto}, {Vernieri}, {Walker}, {Weinberg}, {Young},
  {Anthony-Brumfield}, {Bizyaev}, {Breslauer}, {De Lee}, {Downey}, {Halverson},
  {Huehnerhoff}, {Klaene}, {Leon}, {Long}, {Mahadevan}, {Malanushenko},
  {Nguyen}, {Owen}, {S{\'a}nchez-Gallego}, {Sayres}, {Shane}, {Shectman},
  {Shetrone}, {Skinner}, {Stauffer}, \& {Zhao}}]{wilson2019}
{Wilson}, J.~C., {Hearty}, F.~R., {Skrutskie}, M.~F., {et~al.} 2019, \pasp,
  131, 055001

\bibitem[{{Zasowski} {et~al.}(2013){Zasowski}, {Johnson}, {Frinchaboy},
  {Majewski}, {Nidever}, {Rocha Pinto}, {Girardi}, {Andrews}, {Chojnowski},
  {Cudworth}, {Jackson}, {Munn}, {Skrutskie}, {Beaton}, {Blake}, {Covey},
  {Deshpande}, {Epstein}, {Fabbian}, {Fleming}, {Garcia Hernandez}, {Herrero},
  {Mahadevan}, {M{\'e}sz{\'a}ros}, {Schultheis}, {Sellgren}, {Terrien}, {van
  Saders}, {Allende Prieto}, {Bizyaev}, {Burton}, {Cunha}, {da Costa},
  {Hasselquist}, {Hearty}, {Holtzman}, {Garc{\'\i}a P{\'e}rez}, {Maia},
  {O'Connell}, {O'Donnell}, {Pinsonneault}, {Santiago}, {Schiavon}, {Shetrone},
  {Smith}, \& {Wilson}}]{zasowski2013}
{Zasowski}, G., {Johnson}, J.~A., {Frinchaboy}, P.~M., {et~al.} 2013, \aj, 146,
  81

\bibitem[{{Zasowski} {et~al.}(2017){Zasowski}, {Cohen}, {Chojnowski},
  {Santana}, {Oelkers}, {Andrews}, {Beaton}, {Bender}, {Bird}, {Bovy},
  {Carlberg}, {Covey}, {Cunha}, {Dell'Agli}, {Fleming}, {Frinchaboy},
  {Garc{\'\i}a-Hern{\'a}ndez}, {Harding}, {Holtzman}, {Johnson}, {Kollmeier},
  {Majewski}, {M{\'e}sz{\'a}ros}, {Munn}, {Mu{\~n}oz}, {Ness}, {Nidever},
  {Poleski}, {Rom{\'a}n-Z{\'u}{\~n}iga}, {Shetrone}, {Simon}, {Smith},
  {Sobeck}, {Stringfellow}, {Szigeti{\'a}ros}, {Tayar}, \&
  {Troup}}]{zasowski2017}
{Zasowski}, G., {Cohen}, R.~E., {Chojnowski}, S.~D., {et~al.} 2017, \aj, 154,
  198

\bibitem[{{Zinn} {et~al.}(2019{\natexlab{a}}){Zinn}, {Pinsonneault}, {Huber},
  {Stello}, {Stassun}, \& {Serenelli}}]{joelradius}
{Zinn}, J.~C., {Pinsonneault}, M.~H., {Huber}, D., {et~al.} 2019{\natexlab{a}},
  \apj, 885, 166

\bibitem[{{Zinn} {et~al.}(2019{\natexlab{b}}){Zinn}, {Stello}, {Huber}, \&
  {Sharma}}]{BAM}
{Zinn}, J.~C., {Stello}, D., {Huber}, D., \& {Sharma}, S. 2019{\natexlab{b}},
  \apj, 884, 107

\bibitem[{{Zinn} {et~al.}(2020){Zinn}, {Stello}, {Elsworth}, {Garc{\'\i}a},
  {Kallinger}, {Mathur}, {Mosser}, {Bugnet}, {Jones}, {Hon}, {Sharma},
  {Sch{\"o}nrich}, {Warfield}, {Luger}, {Pinsonneault}, {Johnson}, {Huber},
  {Silva Aguirre}, {Chaplin}, {Davies}, \& {Miglio}}]{k2gapdr2}
{Zinn}, J.~C., {Stello}, D., {Elsworth}, Y., {et~al.} 2020, \apjs, 251, 23

\end{thebibliography}
\bibliographystyle{aasjournal}

\end{document}